\documentclass{aa}
\usepackage{graphicx}
\usepackage{bm}
\usepackage{times}
\graphicspath{{./fig/}{./png/}}

\newcommand{\EQ}{\begin{equation}}
\newcommand{\EE}{\end{equation}}
\newcommand{\EQA}{\begin{eqnarray}}
\newcommand{\EEA}{\end{eqnarray}}

\newcommand{\pd}{\partial}

\newcommand{\DIV}{\vec{\nabla} \cdot }
\newcommand{\CURL}{\vec{\nabla} \times }

\newcommand{\ve}[1]{\boldsymbol{#1}}
\newcommand{\mean}[1]{\overline{#1}}
\newcommand{\meanv}[1]{\overline{\bm #1}}

\newcommand{\nut}{\nu_{\rm t}}

\newcommand{\nuto}{\nu_{\rm t0}}
\newcommand{\urms}{u_{\rm rms}}

\newcommand{\lamh}{\Lambda_{\rm H}}
\newcommand{\lamm}{\Lambda_{\rm M}}
\newcommand{\lamv}{\Lambda_{\rm V}}
\newcommand{\tlamh}{\tilde{\Lambda}_{\rm H}}
\newcommand{\tlamm}{\tilde{\Lambda}_{\rm M}}
\newcommand{\tlamv}{\tilde{\Lambda}_{\rm V}}
\newcommand{\kef}{k_{\rm f}}

\newcommand{\St}{{\rm St}}
\newcommand{\Sh}{{\rm Sh}}
\newcommand{\Co}{{\rm Co}}
\newcommand{\Ma}{{\rm Ma}}
\newcommand{\Rey}{{\rm Re}}

\newcommand{\qxx}{Q_{xx}}
\newcommand{\qyy}{Q_{yy}}
\newcommand{\qzz}{Q_{zz}}
\newcommand{\qxy}{Q_{xy}}
\newcommand{\qxz}{Q_{xz}}
\newcommand{\qyz}{Q_{yz}}
\newcommand{\qij}{Q_{ij}}
\newcommand{\tqxx}{\tilde{Q}_{xx}}
\newcommand{\tqxy}{\tilde{Q}_{xy}}
\newcommand{\tqxz}{\tilde{Q}_{xz}}
\newcommand{\tqyy}{\tilde{Q}_{yy}}
\newcommand{\tqyz}{\tilde{Q}_{yz}}
\newcommand{\tqzz}{\tilde{Q}_{zz}}
\newcommand{\tqij}{\tilde{Q}_{ij}}
\newcommand{\Omx}{\Omega_x}

\def\onethird{{\textstyle{1\over3}}}
\def\twothirds{{\textstyle{2\over3}}}
\def\onehalf{{\textstyle{1\over2}}}

\begin{document}

\authorrunning{Snellman et al.}
\titlerunning{Reynolds stresses from hydrodynamic turbulence with shear and rotation}

   \title{Reynolds stresses from hydrodynamic turbulence with shear and rotation}

   \author{J. E. Snellman
	  \inst{1},
          P. J. K\"apyl\"a
	  \inst{1,2}
          M. J. Korpi
	  \inst{1}
          \and
          A. J. Liljestr\"om
	  \inst{1},
	  }
   \offprints{J.\ E.\ Snellman\\
	  \email{jan.snellman@helsinki.fi}
	  }

   \institute{Observatory, PO Box 14, FI-00014 University of Helsinki, 
              Finland
         \and NORDITA, Roslagstullsbacken 23, SE-10691 Stockholm, Sweden}

   \date{Received 9 June 2009 / Accepted 3 August 2009}

   \abstract{}
   {To study the Reynolds stresses which describe turbulent momentum
     transport from turbulence affected by large-scale shear and
     rotation.}
   {Three-dimensional numerical simulations are used to study
     turbulent transport under the influences of large-scale shear and
     rotation in homogeneous, isotropically forced turbulence. We
     study three cases: one with only shear, and two others where in
     addition to shear, rotation is present. 
     These cases differ by the angle ($0$ or $90\degr$) the rotation
     vector makes with respect to the $z$-direction. 
     Two subsets of runs are performed with both values of $\theta$
     where either rotation or shear is kept constant.
     When only shear is present,
     the off-diagonal stress can be described by turbulent viscosity
     whereas if the system also rotates, nondiffusive contributions 
     ($\Lambda$-effect) to
     the stress can arise. 
     Comparison of
     the direct simulations are made with analytical results from a simple
     closure model.}
   {We find that the turbulent viscosity is of the order of the first
     order smoothing result in the parameter regime studied and that
     for sufficiently large Reynolds numbers the Strouhal number,
     describing the ratio of correlation to turnover times, is roughly
     1.5. This is consistent with the closure model based on the
     minimal tau-approximation which produces a reasonable fit to the
     simulation data for similar Strouhal numbers. In the cases where
     rotation is present, separating the diffusive and nondiffusive
     components of the stress turns out to be challenging but taking
     the results at face value, we can obtain nondiffusive
     contributions of the order of 0.1 times the turbulent viscosity.
     We also find that the simple closure model is able to reproduce
     most of the qualitative features of the numerical results
     provided that the Strouhal number is of the order of unity.}
   {}

   \keywords{   hydrodynamics --
                turbulence --
                accretion disks --
                Sun: rotation --
                stars: rotation                
               }

   \maketitle

%________________________________________________________________

\section{Introduction}
Turbulent angular momentum transport is considered to be of importance
in various astrophysical objects, such as accretion disks (e.g.\
Balbus \& Hawley \cite{BH98}) and convectively unstable layers within
stars (e.g.\ R\"udiger \cite{R89}) where they take part in shaping the
internal rotation profile of the object. Due to the immense numerical
requirements, direct global simulations of these systems are not
yet feasible 
in large quantities, although some simulations are able to
capture many features of, for example, the solar rotation profile (see
e.g.\ Robinson \& Chan \cite{RC01}; Brun \& Toomre \cite{BT02}; Miesch
et al.\ \cite{Mea06,Mea08}). However, in many cases it would be 
desirable to be able
to use simplified, and computationally less demanding, mean-field
models where the effects of small-scale turbulence are accurately
parameterized in some collective way. Although a rich literature of
mean-field models of solar internal rotation exist (e.g.\ Brandenburg
et al.\ \cite{BMT92}, K\"uker et al.\ \cite{KRK93}, Kitchatinov \& 
R\"udiger \cite{KR05}; Rempel
\cite{R05}), many of the models use simple and often untested
parameterizations of the turbulent quantities. 

Parameterizing turbulence entails a closure model for the
turbulent correlations. One of the most often used closures in
astrophysics is the $\alpha$-prescription (Shakura \& Sunyaev
\cite{SS73}), widely used in accretion disk theory, which 
relates the turbulent viscosity to the local gas pressure. This very
simple parameterization allows analytically tractable solutions of
accretion disk structure but suffers from the drawback that important
physics, such as magnetic fields, are not taken into account.
More recently, dynamical turbulent closure models
yielding all the relevant components of the Reynolds and Maxwell
stresses in simplified setups with homogeneous magnetized shear flows
have started to appear (e.g.\ Ogilvie \cite{O03}; Pessah et al.\
\cite{PCP06}).

For these more sophisticated models to be useful, they need to be
validated somehow. The obvious validation method is to compare to
numerical simulations that work in a parameter regime which is as
similar as possible. Comparisons of closure models with numerical
simulations of the magnetorotational instability in periodic local
slab geometry have appeared recently (e.g.\ Pessah et al.\
\cite{PCP08}; Liljestr\"om et al.\ \cite{LKKBL09}) and yield
encouraging results in that the closure models are able to
capture, at least qualitatively, many of the main features of the
numerical simulations. However, these studies concentrate on a
situation where the turbulence is generated via the magnetorotational
instability (Velikhov \cite{V59}; Chandrasekhar \cite{C61}; Balbus \&
Hawley \cite{BH91}), the nonlinear behaviour of which is still not very well
understood (e.g.\ Fromang et al.\ \cite{Fea07}).

In the present paper we avoid these complications and consider only
the simplest possible hydrodynamic case and assume that isotropic 
and homogeneous
background turbulence already exists in the system upon which
large-scale shear and rotation can be imposed. Such turbulence can
be generated by using suitable forcing in the Navier--Stokes equation.
Although flows of this kind are not likely to occur in nature, they
are perfect testbeds for the developement and testing of turbulent
closure models.

The present paper is a continuation to an earlier study (K\"apyl\"a \&
Brandenburg \cite{KB08}, hereafter KB08; see also K\"apyl\"a \&
Brandenburg \cite{KB07}) where simulations of anisotropic homogeneous
turbulence under the influence of rotation were compared to a simple
closure model applying the so-called tau-approximation (hereafter MTA, 
see e.g.\ Blackman
\& Field \cite{BF02,BF03}; Brandenburg et al.\ \cite{BKM04}). In the
present study we adopt an isotropic forcing and add a large-scale
shear flow using the shearing box approximation. Such a setup 
allows us to determine the turbulent
viscosity (see preliminary results in K\"apyl\"a \& Brandenburg
\cite{KB07} and K\"apyl\"a et al.\ \cite{KMB09}). The imposed shear
flow also introduces anisotropy into the turbulence which, under the
influence of rotation, can produce additional non-diffusive Reynolds
stresses (Leprovost \& Kim \cite{LK07,LK08a,LK08b}), which are more
commonly known as the $\Lambda$-effect (Krause \& R\"udiger
\cite{KR74}; R\"udiger \cite{R80},\cite{R89}). We make an effort to
separate the diffusive and nondiffusive contributions from the
numerical data. As in KB08, one of the main goals of the study is to
compare the simulation results to a simple closure model, similar to
that introduced by Ogilvie (\cite{O03}).

The remainder of the paper is organised as follows: the models used 
in the study are
presented in Sect.~\ref{sec:model}, whereas Sects.~\ref{sec:results} and
\ref{sec:conclusions} give the results and conclusions of the study,
respectively.

\section{The model and methods}
\label{sec:model}
\subsection{Basic equations}
We model compressible hydrodynamic turbulence in shearing periodic
cube of size $(2\pi)^3$. The gas obeys an isothermal equation of state
characterized by a constant speed of sound, $c_{\rm s}$. The
continuity and Navier--Stokes equations read
\begin{equation}
\frac{\mathcal{D} \ln \rho}{\mathcal{D} t} = - \vec{\nabla} \cdot \vec{U},
\end{equation}
\begin{equation}
\frac{\mathcal{D} \vec{U}}{\mathcal{D} t} = -SU_x\hat{\bm y} - c_{\rm s}^{2} \vec{\nabla} \ln \rho - 2\, \bm{\Omega} \times \bm{U} + \vec{f}_{\rm visc} + \vec{f}_{\rm force},\label{equ:NS}
\end{equation}
where $\mathcal{D}/\mathcal{D} t=\partial/\partial
t+(\vec{U}+\meanv{U}_0)\cdot\bm{\nabla}$ denotes the advective
derivative and $\meanv{U}_0=(0,Sx,0)$ is the imposed shear
flow. $\vec{U}$ is the velocity, $\rho$ is the density, $\vec{f}_{\rm
  visc}$ is the viscous force, and $\vec{f}_{\rm force}$ is the
forcing function. Compressibility is retained but low Mach number
flows, i.e.\ $\Ma = \urms/c_{\rm s} \approx 0.05-0.1$, where $\urms$ is 
the average root mean square velocity, are
considered. The viscous force is given by
\begin{equation}
\bm{f}_{\rm visc} = \nu \Big(\nabla^2 \bm{U} +\onethird \vec{\nabla} \DIV \bm{U} + 2\, \bm{\mathsf{S}} \cdot \vec{\nabla} \ln \rho \Big)\;,
\end{equation}
where $\nu$ is the kinematic viscosity and
\begin{equation}
\mathsf{S}_{ij} = \onehalf \bigg(\frac{\pd U_i}{\pd x_j} + \frac{\pd U_j}{\pd x_i} \bigg) - \onethird \delta_{ij} \frac{\pd U_k}{\pd x_k},
\end{equation}
is the traceless rate of strain tensor.
The forcing function $\bm{f}_{\rm force}$ is given by
\begin{eqnarray}
\bm{f}(\bm{x},t) = {\rm Re} \{N \bm{f}_{\bm{k}(t)} \exp [i \bm{k}(t)
  \cdot \bm{x} - i \phi(t) ] \}\;,
\end{eqnarray}
where $\bm{x}$ is the position vector, $N = f_0 c_{\rm s} (k c_{\rm
  s}/\delta t)^{1/2}$ is a normalization factor, $f_0$ is the forcing
amplitude, $k = |\bm{k}|$, $\delta t$ is the length of the time step,
and $-\pi < \phi(t) < \pi$ a random delta-correlated phase. The vector
$\bm{f}_{\bm{k}}$ is given by
\begin{eqnarray}
\bm{f}_{\ve{k}} = \frac{\bm{k} \times \hat{\bm{e}}}{\sqrt{\bm{k}^2 - (\bm{k}
    \cdot \hat{\bm{e}})^2}}\;,
\end{eqnarray}
where $\hat{\bm{e}}$ is an arbitrary unit vector. Thus, $\bm{f}_{\bm{k}}$
describes nonhelical transversal waves with $|\bm{f}_{\ve{k}}|^2 = 1$,
where ${\bm k}$ is chosen randomly from a predefined range in the vicinity
of the average wavenumber $\kef/k_1 = 5$ at each time step. Here $k_1$ 
is the wavenumber corresponding to the domain size,
and $\kef$ is the wavenumber of the energy-carrying scale.
The choice $\kef/k_1 = 5$ is somewhat arbitrary but provides a clear
scale separation between the turbulent eddies and the system size.

The numerical simulations were performed with the {\sc Pencil
  Code}\footnote{\texttt{http://www.nordita.org/software/pencil-code/}},
which uses sixth-order accurate finite differences in space, and a
third-order accurate time-stepping scheme (Brandenburg \& Dobler
\cite{BranDobler2002}; Brandenburg
\cite{Brandenburg2003}). Resolutions up to $1024^3$ grid points were
used.

\subsection{Dimensionless units and parameters}

We obtain nondimensional variables by setting
\begin{eqnarray}
c_{\rm s}=k_1=\rho_0=1.
\end{eqnarray}
This means that the units of length, time, and density are
\begin{eqnarray}
[x]=k_1^{-1},\;\;
[t]=(c_{\rm s}k_1)^{-1},\;\;
[\rho]=\rho_0.
\end{eqnarray}
However, in what follows we present the results in explicitly
dimensionless form using the quantities above.

\begin{figure}[t]
\centering
\includegraphics[width=0.5\textwidth]{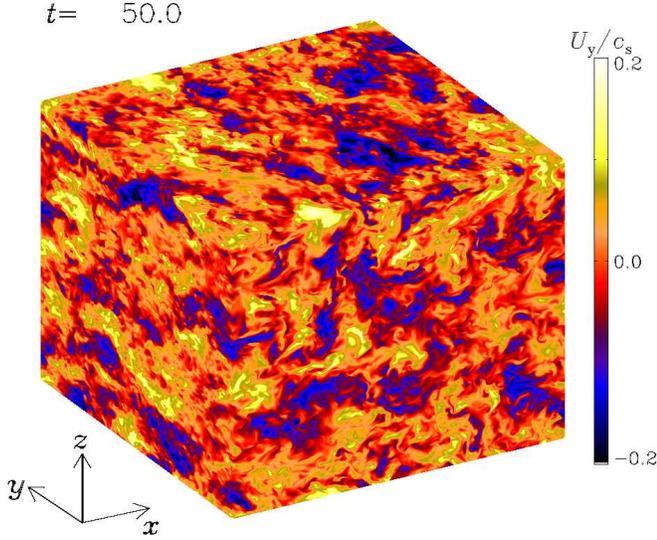}
\caption{Velocity component $U_y$, in the units of the sound speed,
  from the Run~B6 with $\Rey=387$,
  $\Sh=-0.20$, and $\Co=0$, resolution $1024^3$.}
\label{fig:1024a1}
\end{figure}

The strengths of shear, rotation, and viscous effects are measured by the
shear, Coriolis and Reynolds numbers, respectively, based on the 
forcing scale as
\EQA
\Sh = \frac{S}{\urms \kef}, \quad \Co = \frac{2\, \Omega_0}{\urms
  \kef}\;, \quad \label{equ:Coriolis}
{\rm Re} = \frac{\urms}{\nu \kef}\;.
\label{equ:Corioliscomnew}
\EEA
See Fig.~\ref{fig:1024a1} for a typical snapshot from a high
resolution run with $\Rey\approx387$.

\subsection{Coordinate system, averaging, and error estimates}
\label{subsec:coor}
The simulated domain can be thought to represent a small rectangular
portion of a spherical body of gas. We choose $(x,y,z)$ to correspond
to $(\theta,\phi,r)$ of spherical coordinates. With this choice the
rotation vector can be written as
\begin{equation}
\bm{\Omega} = \Omega_0(-\sin \theta,0,\cos \theta)^T,
\label{OmegaDef}
\end{equation}
where $\theta$ is the angle between the rotation axis and the local
vertical ($z$-) direction, i.e.\ the colatitude.
We consider two cases:
$\theta=0$ and $\theta=90\degr$.
We can consider these cases to represent the north pole and the
equator of a rotating star, respectively. However, if an additional a
shear flow of the form $\meanv{U}_0=(0,Sx,0)$ is introduced into the
system there are multiple ways to interpret the physical system that
is described by the model.

Firstly, the case $\theta=0$ can be considered to describe a local
portion of a disk rotating around a central object sufficiently far
away that the effects of curvature can be neglected. Then the rotation
profile of the disk is characterized by $\Omega\propto R^{-q}$ where
$R$ is the radius and $q=-S/\Omega_0$. Now, a Keplerian rotation
profile is obtained for $q=1.5$, a flat profile, such as those 
observed in many
galaxies, is given by $q=1$, and for perfectly rigid rotation we have
$q=0$.

Alternatively, the shear flow can be understood to represent either
radial or latitudinal shear in a convection zone of a star. This
approach has been used by Leprovost \& Kim (\cite{LK07}) who consider
that in the case $\theta=0$, $\meanv{U}^{(0)}$ corresponds to
latitudinal shear near the equator, and in the case $\theta=90\degr$
to the radial shear in near the pole of the star.

Since the turbulence is homogeneous, volume averages are employed and
denoted by overbars. An additional time average over the
statistically saturated state of the simulation is also taken. 
Errors are estimated by dividing the time series into three equally
long parts and computing mean values for each part individually. The
largest departure from the mean value computed for the whole time
series is taken to represent the error.

\subsection{Reynolds stresses from the minimal tau-approximation}
\label{subsec:MTA}
We follow here the same procedure as in KB08 and derive a simple
analytical model for the Reynolds stresses in the case of homogeneous
turbulence under the influences of rotation and shear. Although we
present the model in terms of the minimal tau-approximation (see,
e.g.\ Blackman \& Field \cite{BF02,BF03}; Brandenburg et al.\ \cite{BKM04};
KB08) the closure used here is quite similar to that originally
presented by Ogilvie (\cite{O03}; see also Garaud \& Ogilvie 
\cite{GO05}). A more detailed comparison is
given below.

One of the main purposes of this study is to compare the results from the closure
model with numerical simulations. Since the numerical setup is
homogeneous, it is sufficient to compare the volume averaged data with
a closure model with no spatial extent. In this case, the Navier--Stokes 
equations yield
\begin{equation}
\dot{U}_i  = -SU_x \delta_{iy} -U_j\pd_jU_i  -c_s^2 \partial_i \ln \rho -2 \epsilon_{ilk} \Omega_l U_k + f_i,
\label{dotu}
\end{equation}
where the dot represents time derivative, and $f_i$ describes both 
the viscous force and external forcing.
Now we decompose the velocity as $U_i=\mean{U}_i+u_i$, where $\mean{U}_i$
is the average velocity and $u_i$ the fluctuation. It is straightforward
to derive the equation for $\dot{u}_i$ from Eq.~(\ref{dotu}) which
allows us to derive an evolution equation for the Reynolds stress 
$Q_{ij} = \mean{u_iu_j}$,
\begin{eqnarray}
\dot{Q}_{ij} & = & -S\delta_{yi}Q_{xj}
-S\delta_{yj}Q_{xi} -2 \epsilon_{ilk} \Omega_l Q_{ki} -2 \epsilon_{jlk} \Omega_l Q_{kj}  \nonumber \\
 & & -c_s^2 \overline{ u_i \partial_j 
\ln \rho + u_j \partial_i \ln \rho}\!+\!\mean{\qij\DIV{\bm{u}}}\!+\!\overline{f_i u_j}\!+\!\overline{f_j u_i},
\label{intermed}
\end{eqnarray}
where the overbars denote averaging.
In the minimal tau-approximation closure scheme the nonlinear terms in
Eq.~(\ref{intermed}) are modeled collectively by a relaxation term
\begin{equation}
-c_s^2 \overline{ u_i \partial_j \ln \rho + u_j \partial_i \ln \rho} + \mean{\qij\DIV{\bm{u}}} = -\tau^{-1}Q_{ij},\label{equ:baMTA}
\end{equation}
where $\tau$ is a relaxation time. Terms involving the forcing and
viscosity, $f_i$, can be parameterized in the same spirit with
\begin{equation}
\overline{f_i u_j} + \overline{f_j u_i}=\tau_{\rm f}^{-1}Q_{ij}^{(0)},
\end{equation} 
where $Q_{ij}^{(0)}$ is the equilibrium solution in the absence of 
shear and rotation. Throughout this paper we assume that the
time scale associated with the forcing is equal to
the relaxation time, i.e.\ $\tau=\tau_{\rm f}$. 
We note that there is no compelling theoretical argument to enforce 
this equality but we rather use it for the sake of simplicity.
Thus we arrive at the equation
\begin{eqnarray}
\dot{Q}_{ij} & = & -S\delta_{yi}Q_{xj}
-S\delta_{yj}Q_{xi} -2 \epsilon_{ilk} \Omega_l Q_{ki} -2 \epsilon_{jlk} \Omega_l Q_{kj}  \nonumber \\
 & & \hspace{4cm} -\tau^{-1}(Q_{ij} - Q_{ij}^{(0)}).
\label{finalcl}
\end{eqnarray}
We use the same values of $\Omega$ and $S$ as in the simulations and
take the quantity $Q_{ij}^{(0)}$ from a nonrotating run with
$\Co=\Sh=0$. The free parameter in the model is the relaxation time,
$\tau$, which can be represented in terms of the Strouhal number
\begin{equation}
\St=\tau \urms \kef,\label{equ:St}
\end{equation}
which is the ratio of correlation and turnover times.

\subsubsection{Comparison to the Ogilvie (2003) model}

In the model of Ogilvie (\cite{O03}) the Reynolds stresses in the
hydrodynamical case are given by
\begin{eqnarray}
\dot{Q}_{ij} + \mbox{linear terms} &=& \nonumber \\
&& \hspace{-2cm} C_1 \sqrt{Q} L^{-1}\qij + C_2 \sqrt{Q} L^{-1} (\qij - \onethird Q \delta_{ij}),\label{equ:O03}
\end{eqnarray}
where $C_i$ are dimensionless parameters of the order of unity, $Q$ is
the trace of the Reynolds tensor and $L$ is a lenght scale, e.g.\ the 
system size. The first 
term on the rhs can be identified as a relaxation term similar to that 
used in the minimal tau-approximation
\begin{equation}
\tau=\frac{L}{C_1 \sqrt{Q}} = \frac{2\pi \kef}{C_1 k_1} (\urms \kef)^{-1},
\end{equation}
where we have used $\sqrt{Q}=\urms$, and $L=2\pi/k_1$. Comparing with
Eq.~(\ref{equ:St}), we see that $\St=2\pi\kef
(C_1k_1)^{-1}=10\pi\,C_1^{-1}$ for our value of $\kef/k_1=5$.

Our Eq.~(\ref{finalcl}) also lacks the term $C_2 \sqrt{Q}
L^{-1}(Q_{ij}-\frac{1}{3} Q \delta_{ij})$ which appears in the model
of Ogilvie. This term describes isotropization of the turbulence, and a
similar term can be added to the minimal tau-approximation in the form
$\tau_{\rm I}^{-1}(Q_{ij}-\frac{1}{3} Q \delta_{ij})$, where
$\tau_{\rm I}$ could have a value unequal to $\tau$. However, after
experimenting with such a term, we did not find substantially better
agreement with simulation results. Thus, in order to keep our model as
simple as possible with the minimum amount of free parameters, also
this term is neglected in our analysis.

We note that if the saturated solutions of Eqs.~(\ref{finalcl}) and
(\ref{equ:O03}) are independent of time, the two models yield the same
results provided that $C_2$ is zero. However, if $\dot{Q}\neq0$ in the
saturated state, e.g.\ if the solution is oscillatory, then
differences will occur because the relaxation time in our model is
based on a constant $\urms$ and not on the $Q$ that emerges as a
results of time integration of the model itself. We also stress that
in the original work of Ogilvie (\cite{O03}), the turbulence \emph{due to}
the shear flow, rotation, and magnetic fields was studied. In our case, 
however, the
turbulence \emph{pre-exists} due to the external forcing and we mainly study
the effects of shear and rotation on this background turbulence.

\section{Results}
\label{sec:results}
We study the Reynolds stresses from three different systems: one where
only shear flow is present and two others where nonzero rotation is
present with either $\theta=0$ or $\theta=90\degr$. For the first
system we perform four different sets of simulations (see
Table~\ref{tab:simsets} for a summary of the different sets of runs) 
where we either fix $\Sh$ (Sets~A and B) or
$\Rey$ (Sets~C and D). In the setups with rotation we perform two
sets of calculations for both values of $\theta$ where we fix either
$\Sh$ (Sets~E and G) or $\Co$ (Sets~F and H). In Sets~E to H we vary
the parameter $q$ in the range $-5<q<1.9$. More detailed descriptions
of the runs can be found in Tables~\ref{tab:onlyshear} to
\ref{tab:setGH}. Furthermore, we compute the stresses from corresponding
closure models making use of the stationary solution of
Eq.~(\ref{finalcl}) given in Appendix~\ref{sec:appendix}.

\begin{table}
  \centering
  \caption[]{Summary of the different sets of simulations. The values 
    of $\Rey$, $\Co$, and $\Sh$ are given in terms of $\urms$ from a 
    run with $\Co=\Sh=0$.}
  \vspace{-0.5cm}
  \label{tab:simsets}
  $$
  \begin{array}{p{0.03\linewidth}ccccccccrrr}
    \hline
    \noalign{\smallskip}
    Set   & \Rey  & \Co & -\Sh & \theta & q \\
    \noalign{\smallskip}
    \hline
    \noalign{\smallskip}
    A    & 5\ldots345 & 0 & \approx0.1 & - & - \\
    B    & 8\ldots387 & 0 & \approx0.2 & - & - \\
    C    & \approx27  & 0 & 0.01\ldots0.24 & - & - \\
    D    & \approx75  & 0 & 0.01\ldots0.23 & - & - \\
    \hline
    \noalign{\smallskip}
    E    & \approx24  & -1.27\ldots1.28 & \approx0.16     & 0 & -5\ldots1.9 \\
    F    & \approx24  & \approx0.32     & -0.29\ldots0.64 & 0 & -5\ldots1.9 \\
    \hline
    \noalign{\smallskip}
    G    & \approx24  & -1.27\ldots1.28 & \approx0.16     & 90\degr & -5\ldots1.9 \\
    H    & \approx24  & \approx0.32     & -0.29\ldots0.64 & 90\degr & -5\ldots1.9 \\
    \hline
   \end{array}
   $$ 
\end{table}

\onltab{1}{
   \begin{table*}
   \centering
   \caption[]{Summary of the runs with shear flow only. Here 
     $\tilde{Q}_{ij}=\qij/\urms^2$, and brackets around a quantity signify 
     that the number is not statistically significant.}
   \vspace{-0.5cm}
      \label{tab:onlyshear}
     $$
         \begin{array}{p{0.03\linewidth}ccccccccrrr}
           \hline
           \noalign{\smallskip}
           Run      & $Grid$ & {\rm Re}  & \Co & \Sh & \urms & \tilde{Q}_{xx} & \tilde{Q}_{yy} & \tilde{Q}_{zz} & \tilde{Q}_{xy} & \tilde{Q}_{xz} & \tilde{Q}_{yz} \\
           \noalign{\smallskip}
           \hline
           \noalign{\smallskip}
           A1 &   64^3 &   5 & 0 & -0.18 & 0.054 & 0.295 & 0.404 & 0.302 & 0.075 &(-0.000) &(-0.001) \\ % 64a9
           A2 &  128^3 &  14 & 0 & -0.14 & 0.069 & 0.304 & 0.378 & 0.319 & 0.084 &(-0.001) &( 0.002) \\ % 128a3
           A3 &  256^3 &  31 & 0 & -0.12 & 0.079 & 0.313 & 0.368 & 0.320 & 0.083 &(-0.002) &(-0.001) \\ % 256a1
           A4 &  256^3 &  80 & 0 & -0.12 & 0.083 & 0.331 & 0.342 & 0.328 & 0.069 &(-0.001) &(-0.001) \\ % 256a4
           A5 &  512^3 & 164 & 0 & -0.12 & 0.084 & 0.318 & 0.349 & 0.333 & 0.058 &(-0.003) &(-0.003) \\ % 512a1
           A6 & 1024^3 & 345 & 0 & -0.11 & 0.088 & 0.320 & 0.343 & 0.337 & 0.054 &(-0.001) &(-0.003) \\ % 1024a2
           \hline
           \noalign{\smallskip}
           B1 &   64^3 &   8 & 0 & -0.25 & 0.080 & 0.198 & 0.599 & 0.238 & 0.146 & (0.002) & (0.001) \\ % 64a2
           B2 &  128^3 &  22 & 0 & -0.18 & 0.112 & 0.235 & 0.491 & 0.288 & 0.140 &(-0.001) &(-0.000) \\ % 128a
           B3 &  256^3 &  44 & 0 & -0.18 & 0.112 & 0.262 & 0.440 & 0.303 & 0.124 &(-0.002) & (0.000) \\ % 256a
           B4 &  512^3 &  99 & 0 & -0.19 & 0.101 & 0.282 & 0.425 & 0.318 & 0.110 & (0.001) & (0.001) \\ % 512a
           B5 & 1024^3 & 196 & 0 & -0.20 & 0.100 & 0.300 & 0.377 & 0.323 & 0.101 &(-0.003) &(-0.001) \\ % 1024a
           B6 & 1024^3 & 387 & 0 & -0.20 & 0.099 & 0.299 & 0.375 & 0.326 & 0.095 &(-0.004) &(-0.005) \\ % 1024a1
          \hline
           \noalign{\smallskip}
           C1 &  128^3 &  27 & 0 & -0.01 & 0.069 & 0.335 & 0.334 & 0.332 & 0.010 &(-0.003) &(-0.002) \\ % 128aa10
           C2 &  128^3 &  27 & 0 & -0.03 & 0.069 & 0.334 & 0.332 & 0.335 & 0.014 &(-0.001) & (0.001) \\ % 128aa9
           C3 &  128^3 &  28 & 0 & -0.06 & 0.070 & 0.335 & 0.332 & 0.333 & 0.031 &(-0.002) & (0.000) \\ % 128aa8
           C4 &  128^3 &  31 & 0 & -0.13 & 0.078 & 0.317 & 0.362 & 0.321 & 0.084 &(-0.001) &(-0.000) \\ % 128aa7
           C5 &  128^3 &  39 & 0 & -0.20 & 0.099 & 0.282 & 0.392 & 0.330 & 0.125 & (0.002) &(-0.002) \\ % 128aa6
           C6 &  128^3 &  48 & 0 & -0.24 & 0.122 & 0.259 & 0.396 & 0.347 & 0.137 & (0.000) & (0.010) \\ % 128aa11
          \hline
           \noalign{\smallskip}
           D1 &  256^3 &  75 & 0 & -0.01 & 0.076 & 0.331 & 0.334 & 0.335 & 0.009 & (0.002) & (0.002) \\ % 256aa10
           D2 &  256^3 &  75 & 0 & -0.03 & 0.076 & 0.335 & 0.329 & 0.336 & 0.015 & (0.001) & (0.000) \\ % 256aa9
           D3 &  256^3 &  76 & 0 & -0.05 & 0.077 & 0.329 & 0.334 & 0.337 & 0.029 &(-0.001) &(-0.000) \\ % 256aa8
           D4 &  256^3 &  81 & 0 & -0.12 & 0.083 & 0.331 & 0.342 & 0.328 & 0.069 & (0.001) &(-0.000) \\ % 256aa7
           D5 &  256^3 &  99 & 0 & -0.19 & 0.101 & 0.297 & 0.373 & 0.331 & 0.108 &(-0.000) & (0.001) \\ % 256aa6
           D5 &  256^3 & 123 & 0 & -0.23 & 0.126 & 0.267 & 0.417 & 0.317 & 0.129 & (0.002) & (0.010) \\ % 256aa11
          \hline
         \end{array}
     $$ 
   \end{table*}
}

\subsection{Case $\Omega=0$: turbulent viscosity}

Consider first the case where rotation is absent and the only
large-scale flow is the imposed shear $\meanv{U} = (0,xS,0)$ with
isotropic background turbulence. The Reynolds stress generated
by the shear can be represented by the expression
\begin{equation}
\qij = -\nut (\mean{U}_{i,j} + \mean{U}_{j,i}) = -\nut S \delta_{ix} \delta_{jy}, \label{equ:nut}
\end{equation}
where $\nut$ is the turbulent viscosity. The expression
(\ref{equ:nut}) can be considered to be valid for weak shear. On the
other hand, we can estimate the turbulent viscosity from
\begin{equation}
\nuto = \onethird \tau \mean{\bm{u}}^2,\label{equ:nutfosa}
\end{equation}
where $\tau$ is the correlation time and $\overline{\bm u^2}$ is the
average turbulent velocity squared. Equation~(\ref{equ:nutfosa}) is
the same as the one that can be derived for the magnetic diffusivity
using first order smoothing approximation (e.g.\ Krause \& R\"adler
\cite{KR80}). Here we have assumed a turbulent magnetic Prandtl number
of unity to arrive at Eq.~(\ref{equ:nutfosa}). This formulation is
supported by numerical results although analytical studies often yield
somewhat different results (e.g.\ Yousef et al. \cite{Yea03} and
references therein).

The correlation time can be
related to the turnover time of the turbulence via the Strouhal
number, Eq.~(\ref{equ:St}).
If we assume that $\St=1$, as is suggested by numerical turbulence
models (e.g.\ Brandenburg \& Subramanian \cite{BS05}, \cite{BS07})
similar to ours, the turbulent viscosity is given by
\begin{equation}
\nuto = \onethird \urms \kef^{-1}.\label{equ:nut0}
\end{equation}
In what follows we use $\nuto$ as the normalization for the turbulent
viscosity.
However, later on we allow $\St\neq1$ and write $\nut=\St \nuto$ in
order to obtain an independent estimate of $\St$ from the numerical 
simulations.

\subsubsection{Simulation results}
\label{sec:simnoom}
We find that in the non-rotating case it is difficult to avoid
large-scale vorticity generation (see also K\"apyl\"a et al.\
\cite{KMB09}). This phenomenon is likely to be related to a vorticity
dynamo discussed in the analytical studies of Elperin et al.\
(\cite{EKR03,EGKR07}). Similar large-scale structures have been
observed earlier in the numerical works of Yousef et al.\
(\cite{Y08a,Y08b}) using an independent method. The
problem becomes increasingly worse as the shear is increased. Thus we
need to limit the range of $\Sh$ and carefully analyze the data so that
the effect of the vorticity dynamo on the turbulence is
minimized (see the related discussion in Mitra et al.\ \cite{MKTB09}). 
What this means is that we limit the time averaging
to a range where turbulence is established but where
the large-scale flow is still weak in comparison to the
turbulent rms-velocity.
We stress that the vorticity dynamo occurs only if turbulence already
exists in the system, i.e.\ it does not arise if the forcing is turned
off in the simulations.

\begin{figure}[t]
\centering
\includegraphics[width=0.5\textwidth]{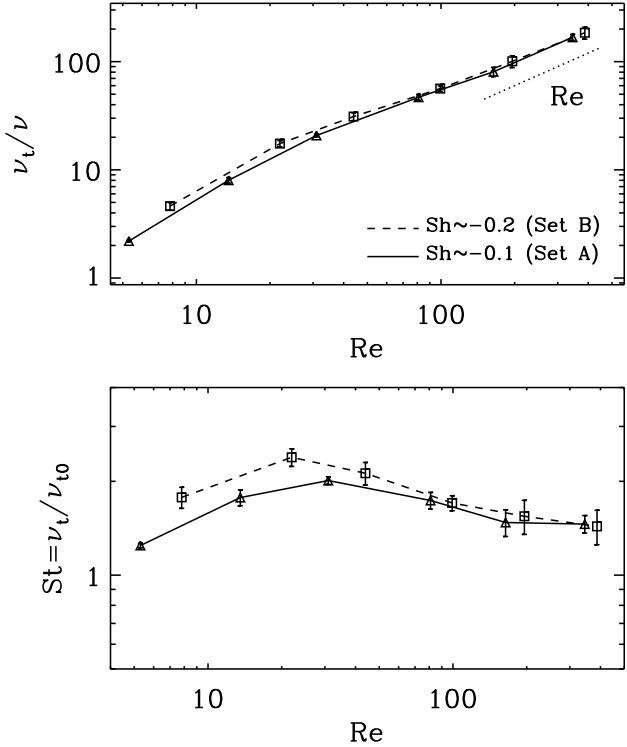}
\caption{Upper panel: turbulent viscosity divided by the molecular
  viscosity for the simulation Sets~A (solid line) and B (dashed line)
  as a function of the Reynolds number. The dotted line, proportional 
  to $\Rey$, is shown for reference. Lower panel:
  the Strouhal number, $\St=\nut/\nuto$, for the same sets of
  simulations as in the upper panel.}
\label{fig:turbvisc}
\end{figure}

The only nonzero off-diagonal component in the simulations is now
$\qxy$ that can be interpreted in terms of turbulent viscosity.
This is also consistent with symmetry arguments.
Once a suitable averaging interval has been found, the turbulent
viscosity is obtained from Eq.~(\ref{equ:nut}). If the first order
smoothing result, Eq.~(\ref{equ:nut0}) is valid, we should expect the ratio of turbulent to
molecular viscosity to be proportional to the Reynolds number,
\begin{equation}
\nut/\nu \propto \Rey.
\end{equation}
We find this scaling to be valid for large
enough $\Rey$ in the simulation results of Sets~A and B, see the upper
panel of Fig.~\ref{fig:turbvisc}. The vorticity generation is more
vigorous for smaller Reynolds numbers which can explain the deviating values
for small $\Rey$. The vorticity dynamo exists at least up to
$\Rey\approx100$ (K\"apyl\"a et al.\ \cite{KMB09}), but the
growth rate seems to decrease somewhat as $\Rey$ is increased.
We also find that the turbulence becomes more isotropic as $\Rey$
increases (cf.\ Table~\ref{tab:onlyshear}) and conjecture that this
behaviour is also linked to the less efficient large-scale vorticity
generation.

Furthermore, we can estimate the Strouhal number by comparing the
simulation results for $\nut$ and the FOSA estimate,
Eq.~(\ref{equ:nut0}), see the lower panel of
Fig.~\ref{fig:turbvisc}. We find that for large Reynolds numbers,
$\nut/\nuto \approx 1.5$, indicating that $\St\approx1.5$. This value
is in accordance with earlier numerical studies of passive scalar
transport (Brandenburg et al.\ \cite{BKM04}) and nondiffusive Reynolds
stresses (K\"apyl\"a \& Brandenburg \cite{KB08}).

\begin{figure}[t]
\centering
\includegraphics[width=0.45\textwidth]{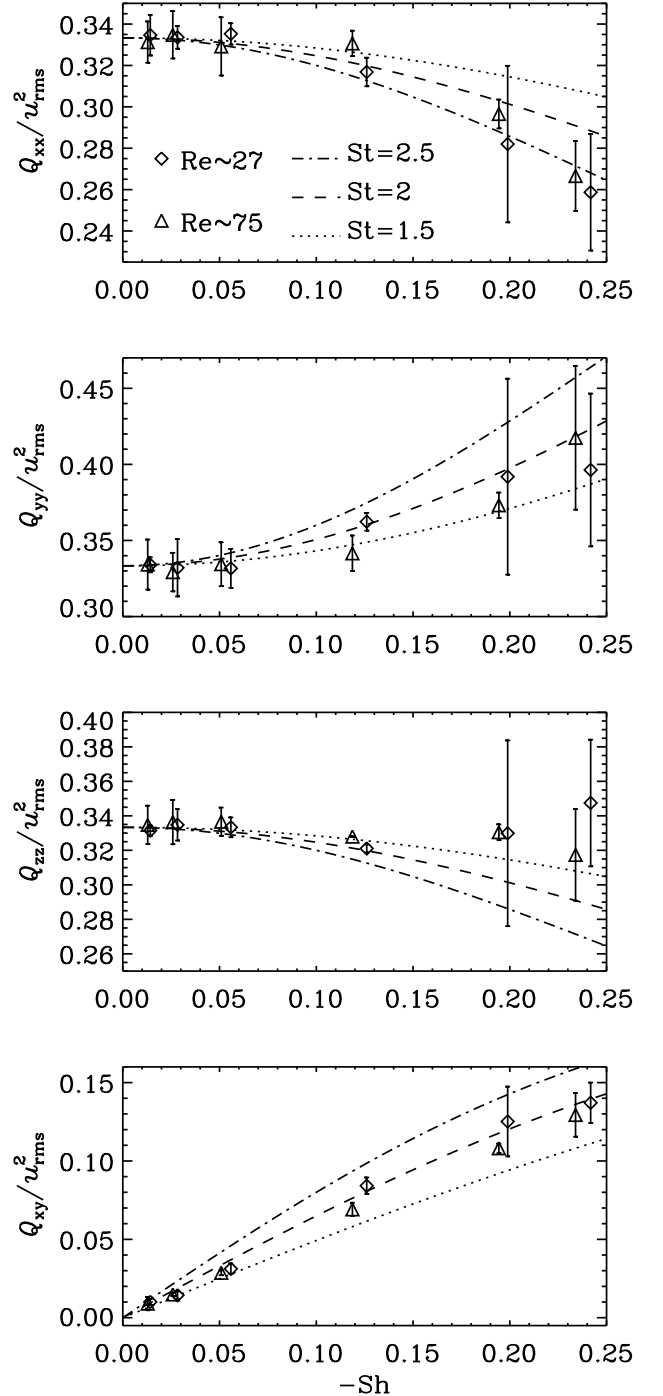}
\caption{From top to bottom: Reynolds stress components $\qxx$,
  $\qyy$, $\qzz$, and $\qxy$, normalised by $\urms^2$, as
  functions of the shear parameter for simulation Sets~C and D with 
  Reynolds numbers, based on $\urms$ from a run with $\Co=\Sh=0$, 
  of $\approx27$ (diamonds) and $\approx75$ (triangles), respectively. The
  curves in each panel show the results of the MTA-closure with three
  Strouhal numbers. The linestyles are as indicated in the legend in the top panel.}
\label{fig:p128aa}
\end{figure}

The shear dependence of the non-zero components of the normalized Reynolds
stresses, $\tqij\equiv\qij/\urms^2$, from Sets~C and D is shown in 
Fig.~\ref{fig:p128aa} (see also
Table~\ref{tab:onlyshear}). We find that $\tqxx$ decreases and
$\tqyy$ increases as a function of $\Sh$, although the latter trend
is barely significant due to the large error bars. $\tqzz$, on the other hand,
seems to increase slightly for strong shear but the error bars are again so
large that this trend is not statistically significant. For
$-\Sh\ga0.2$ the errors are generally quite large due to the short
averaging interval that we are forced to use due to the vorticity
generation.

We can rewrite
Eq.~(\ref{equ:nut}) in terms of $\Sh$ and $\St$ by substituting
$\nuto=\onethird \St \urms \kef^{-1}$ and using the definition of
$\Sh$ to obtain
\begin{equation}
\tilde{Q}_{xy} = -\onethird \St \Sh.
\end{equation}
This relation is valid for weak shear and is also borne out of the
closure model (see 
Appendix~\ref{sucsec:appos}). We find that the stress component
$\tqxy$ in the Sets~C and D is approximately consistent with a
relation $-\tqxy \propto (0.5 \ldots 0.67) \Sh$, suggesting
that $\St\approx1.5\ldots2$ (see the bottom panel of
Fig.~\ref{fig:p128aa} and the next section). The Reynolds number in
the Sets~C and D varies between 27 and 123 so the results for the
Strouhal number are consistent with the corresponding data in
Fig.~\ref{fig:turbvisc}.

\subsubsection{Closure model results}

We now turn to the simple closure model that was introduced in
Sect.~\ref{subsec:MTA}. We compare the stationary solutions of the
MTA-model with the time and volume averaged simulation data.
The closure model predicts that the absolute
values of $\qxx$ and $\qzz$ remain constant as functions of $\Sh$ (see
Appendix~\ref{sec:appendix} for more details)
\begin{equation}
\qxx=\qxx^{(0)}, \quad \qzz=\qzz^{(0)}.
\end{equation}
However, since $\qyy$, and thus $Q=\urms^2$, can vary, the normalised
data shows a decreasing trend for $\qxx$ and $\qzz$ as a function of
shear. The MTA-model also shows that $\qxy$ is linearly proportional
to the shear parameter for weak shear whereas the other two
off-diagonal components are zero as expected from symmetry
cosiderations.

We find that the closure model is in qualititive agreement with the
numerical results for $\tqxx$, $\tqyy$, and $\tqxy$ for all values of
$\Sh$, see Fig.~\ref{fig:p128aa}. For $\tqzz$ the numerical data shows
large scatter for increasing values of $|\Sh|$, although the points around
$-\Sh\approx0.12$ are consistent with a declining trend. 
Considering the other components, the best fit to the simulation data
for $\tqyy$ and $\tqxy$ is obtained if $\St\approx1.5\ldots2$ is used
in the MTA-model (see the dotted and dashed lines in
Fig.~\ref{fig:p128aa}), whereas a somewhat larger $\St$ fits the
results of $\tqxx$ best. Especially the off-diagonal component $\tqxy$
is quite well reproduced with our simple model.

\onltab{2}{
   \begin{table*}
   \centering
   \caption[]{Summary of the runs in the Sets~E and F with $\theta=0$.}
   \vspace{-0.5cm}
      \label{tab:setE}
     $$
         \begin{array}{p{0.03\linewidth}ccrrrccccrrr}
           \hline
           \noalign{\smallskip}
           Run      & $Grid$ & {\rm Re}  & q & \Co & \Sh & \urms & \tilde{Q}_{xx} & \tilde{Q}_{yy} & \tilde{Q}_{zz} & \tilde{Q}_{xy} & \tilde{Q}_{xz} & \tilde{Q}_{yz} \\
           \noalign{\smallskip}
           \hline
           \noalign{\smallskip}
           E1 &  128^3 &  24 &-5.00 & -0.06 & -0.16 & 0.062 & 0.324 & 0.348 & 0.328 & 0.043 & (0.000) &(-0.000) \\ % S005cst/128bb
           E2 &  128^3 &  24 &-4.00 & -0.08 & -0.16 & 0.062 & 0.323 & 0.348 & 0.330 & 0.043 & (0.000) & (0.000) \\ % 128bb1
           E3 &  128^3 &  24 &-3.00 & -0.11 & -0.16 & 0.062 & 0.321 & 0.350 & 0.329 & 0.041 & (0.001) & (0.001) \\ % 128bb2
           E4 &  128^3 &  24 &-2.50 & -0.13 & -0.16 & 0.061 & 0.320 & 0.350 & 0.330 & 0.039 & (0.000) & (0.000) \\ % 128bb3
           E5 &  128^3 &  24 &-2.00 & -0.16 & -0.16 & 0.061 & 0.318 & 0.352 & 0.330 & 0.036 & (0.001) &(-0.000) \\ % 128bb4
           E6 &  128^3 &  24 &-1.50 & -0.21 & -0.16 & 0.061 & 0.317 & 0.353 & 0.330 & 0.033 & (0.000) &(-0.001) \\ % 128bb5
           E7 &  128^3 &  24 &-1.00 & -0.32 & -0.16 & 0.061 & 0.315 & 0.353 & 0.332 & 0.027 &(-0.000) &(-0.000) \\ % 128bb6
           E8 &  128^3 &  24 &-0.50 & -0.64 & -0.16 & 0.061 & 0.315 & 0.351 & 0.335 & 0.014 & (0.001) &(-0.001) \\ % 128bb7
           E9 &  128^3 &  24 &-0.25 & -1.27 & -0.16 & 0.062 & 0.318 & 0.341 & 0.341 & 0.005 & (0.001) &(-0.001) \\ % 128bb8
          E10 &  128^3 &  24 & 0.25 &  1.28 & -0.16 & 0.061 & 0.352 & 0.310 & 0.337 & 0.003 &(-0.001) &(-0.000) \\ % 128bb9
          E11 &  128^3 &  24 & 0.50 &  0.65 & -0.16 & 0.061 & 0.361 & 0.308 & 0.331 & 0.015 &(-0.001) &(-0.001) \\ % 128bb10
          E12 &  128^3 &  24 & 0.75 &  0.43 & -0.16 & 0.061 & 0.361 & 0.309 & 0.330 & 0.026 &(-0.001) &(-0.000) \\ % 128bb11
          E13 &  128^3 &  24 & 1.00 &  0.32 & -0.16 & 0.061 & 0.360 & 0.311 & 0.329 & 0.032 &(-0.001) &(-0.001) \\ % 128bb12
          E14 &  128^3 &  24 & 1.25 &  0.26 & -0.16 & 0.061 & 0.358 & 0.313 & 0.329 & 0.037 &(-0.001) &(-0.000) \\ % 128bb13
          F15 &  128^3 &  24 & 1.50 &  0.21 & -0.16 & 0.061 & 0.357 & 0.315 & 0.328 & 0.040 & (0.000) &(-0.001) \\ % 128bb14
          E16 &  128^3 &  24 & 1.75 &  0.18 & -0.16 & 0.062 & 0.355 & 0.317 & 0.328 & 0.043 &(-0.001) &(-0.001) \\ % 128bb15
          E17 &  128^3 &  24 & 1.90 &  0.17 & -0.16 & 0.062 & 0.355 & 0.318 & 0.327 & 0.044 &(-0.001) &(-0.001) \\ % 128bb16
           \hline
           \noalign{\smallskip}
           F1 &  128^3 &  30 &-5.00 & 0.25 &  0.64 & 0.077 & 0.230 & 0.431 & 0.340 &-0.075 &(-0.001) & (0.000) \\ % 128bb
           F2 &  128^3 &  28 &-4.00 & 0.27 &  0.55 & 0.072 & 0.253 & 0.407 & 0.340 &-0.068 &(-0.001) & (0.001) \\ % 128bb1
           F3 &  128^3 &  26 &-3.00 & 0.29 &  0.44 & 0.067 & 0.276 & 0.388 & 0.336 &-0.058 &(-0.001) & (0.002) \\ % 128bb2
           F4 &  128^3 &  26 &-2.50 & 0.30 &  0.37 & 0.066 & 0.287 & 0.381 & 0.333 &-0.053 &(-0.001) & (0.002) \\ % 128bb3
           F5 &  128^3 &  25 &-2.00 & 0.31 &  0.31 & 0.064 & 0.297 & 0.370 & 0.333 &-0.046 &(-0.002) & (0.002) \\ % 128bb4
           F6 &  128^3 &  25 &-1.50 & 0.31 &  0.24 & 0.063 & 0.306 & 0.363 & 0.330 &-0.038 &(-0.001) & (0.003) \\ % 128bb5
           F7 &  128^3 &  24 &-1.00 & 0.32 &  0.16 & 0.061 & 0.314 & 0.355 & 0.331 &-0.027 &(-0.002) & (0.004) \\ % 128bb6
           F8 &  128^3 &  24 &-0.50 & 0.33 &  0.08 & 0.060 & 0.325 & 0.344 & 0.331 &-0.013 &(-0.002) & (0.002) \\ % 128bb7
           F9 &  128^3 &  24 &-0.25 & 0.33 &  0.04 & 0.060 & 0.328 & 0.341 & 0.331 &-0.007 &(-0.002) & (0.003) \\ % 128bb8
          F10 &  128^3 &  24 & 0.25 & 0.33 & -0.04 & 0.060 & 0.342 & 0.328 & 0.329 & 0.007 &(-0.002) & (0.002) \\ % 128bb9
          F11 &  128^3 &  24 & 0.50 & 0.33 & -0.08 & 0.060 & 0.348 & 0.321 & 0.330 & 0.016 &(-0.001) & (0.002) \\ % 128bb10
          F12 &  128^3 &  24 & 0.75 & 0.32 & -0.12 & 0.061 & 0.352 & 0.317 & 0.331 & 0.023 &(-0.000) & (0.002) \\ % 128bb11
          F13 &  128^3 &  24 & 1.00 & 0.32 & -0.16 & 0.061 & 0.360 & 0.310 & 0.325 & 0.033 &(-0.002) & (0.001) \\ % 128bb12
          F14 &  128^3 &  24 & 1.25 & 0.32 & -0.20 & 0.062 & 0.367 & 0.308 & 0.323 & 0.040 & (0.001) & (0.002) \\ % 128bb13
          F15 &  128^3 &  25 & 1.50 & 0.31 & -0.23 & 0.063 & 0.374 & 0.303 & 0.322 & 0.051 &(-0.000) & (0.002) \\ % 128bb14
          F16 &  128^3 &  25 & 1.75 & 0.31 & -0.27 & 0.064 & 0.381 & 0.298 & 0.319 & 0.060 &(-0.001) &(-0.001) \\ % 128bb15
          F17 &  128^3 &  25 & 1.90 & 0.30 & -0.29 & 0.065 & 0.389 & 0.292 & 0.319 & 0.066 & (0.001) & (0.003) \\ % 128bb16
          \hline
         \end{array}
     $$ 
   \end{table*}
}

\subsection{Case $\Omega\neq0$, $\theta=0$}

When adding shear, we consider two distinct lines in the parameter 
space: in
simulation Set~E we keep $\Sh$ constant and vary $\Co$, whereas in
Set~F the rotation is kept constant and the shear is varied (see also 
Table~\ref{tab:setE}). The range
in which these parameters is varied is given by $-5\le q\le1.9$ where
the parameter $q=-S/\Omega_0$ can be considered to describe the shear
in a differentially rotating disk (see Sect.~\ref{subsec:coor}).
Whilst the cases $q=1.5$ and $q=1$ can be thought to represent
Keplerian and galactic disks, respectively, the regime $q<1$ is not
likely to occur in the bulk of the disk in any system, but such
configurations can occur in convectively unstable parts of stellar
interiors, such as the solar convection zone.
Cases $q>2$ are
Rayleigh unstable (see Appendix~\ref{sec:vortdyn}) and lead to a
large-scale instability of the shear flow so simulations in this
parameter regime are not considered here.

\subsubsection{Simulation results}

\begin{figure}[t]
\centering
\includegraphics[width=0.45\textwidth]{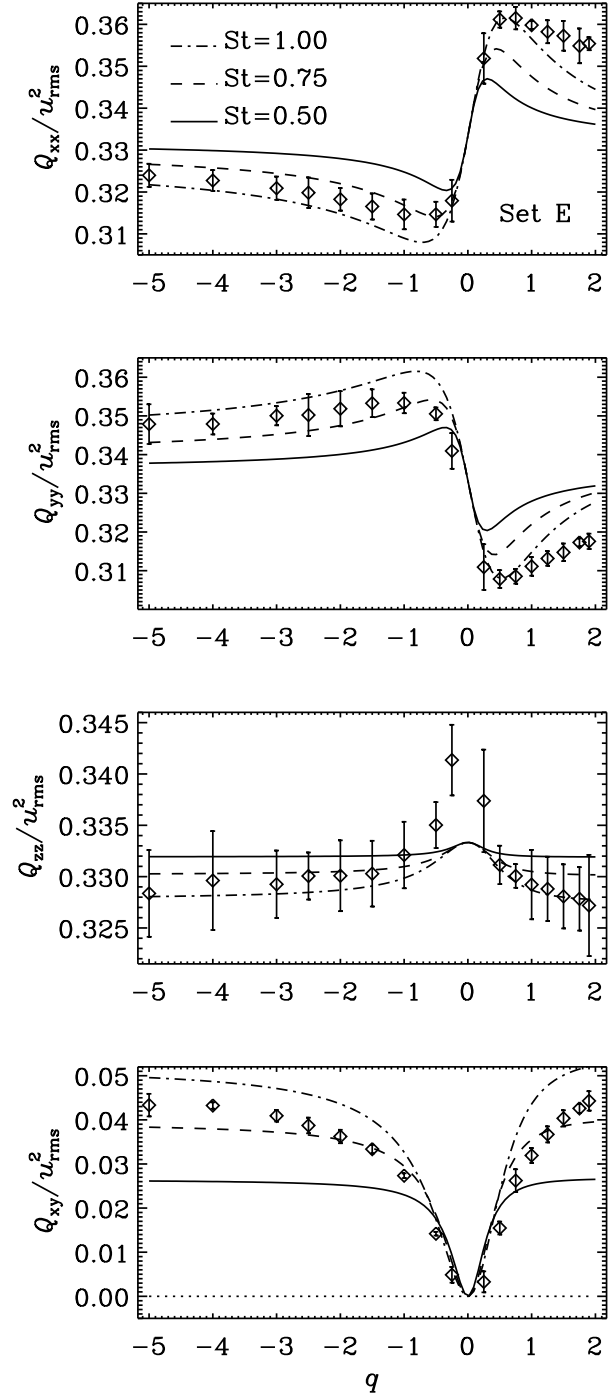}
\caption{Same as Fig.~\ref{fig:p128aa} but for Set~E, and different
  Strouhal numbers as indicated by the legend in the uppermost panel.}
\label{fig:p128bb_Sconst}
\end{figure}

In Set~E we fix the shear flow, measured by $\Sh$, and vary the 
rotation rate. In this case the value of $\Sh\approx-0.16$ is very 
close to being constant in all runs. The simulation results for the
nonzero components of the stress are shown in
Fig.~\ref{fig:p128bb_Sconst}. We find that the components $\tqxx$ and
$\tqyy$ behave very similarly to each other but with opposite trends
as functions of $q$, i.e.\ the sum of the two is roughly constant.
The quantity $\tqzz$ varies much less than the two other diagonal 
components, and is consistent with a constant value as a function of 
$q$ within the error 
bars in all cases apart from three points in the range $|q|\le0.5$. 
The off-diagonal
stress $\tqxy$ shows almost a symmetric profile with respect to $q=0$,
with somewhat steeper rise on the positive side. We note that since
$\Sh\approx\mbox{const}<0$, the very small values of $\tqxy$ around
$q=0$ indicate that the more rapid rotation there either quenches the
turbulent viscosity or that additional nondiffusive contributions with 
the opposite sign are present.

\begin{figure}[t]
\centering
\includegraphics[width=0.45\textwidth]{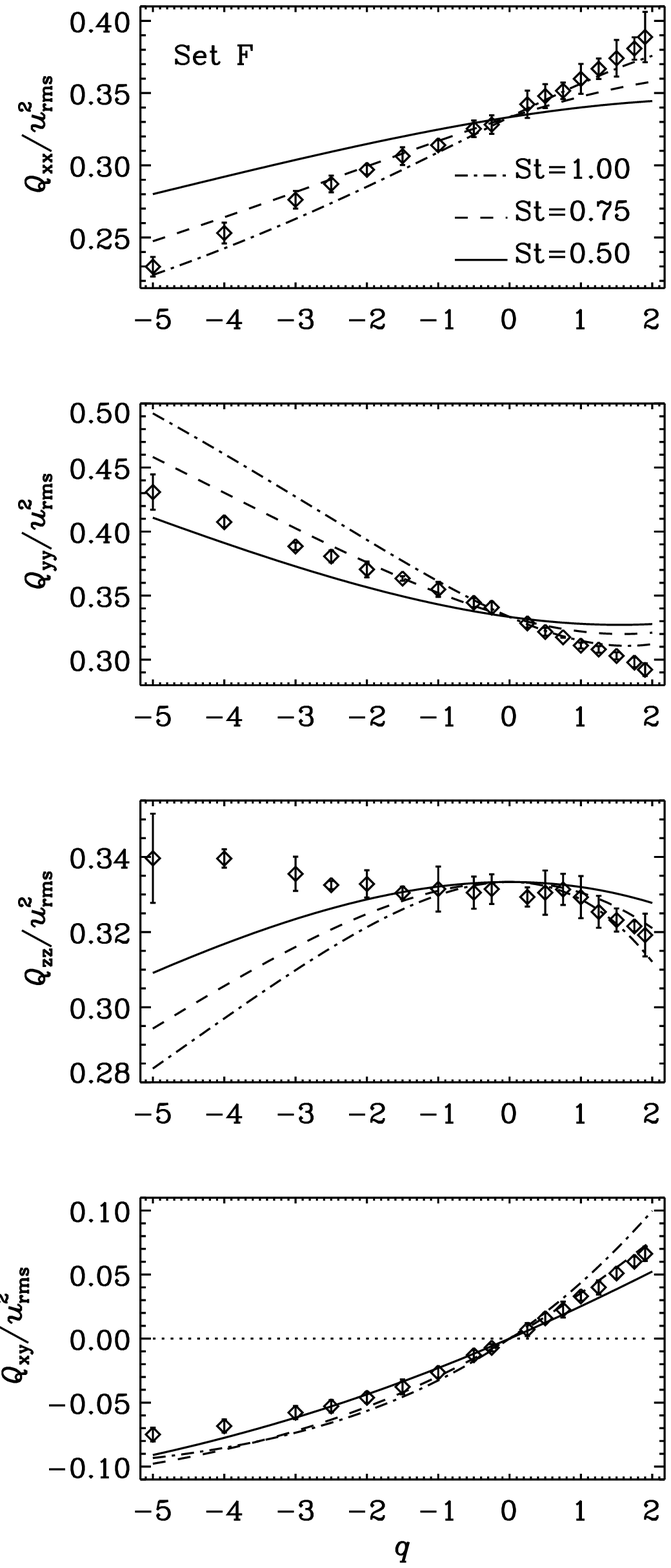}
\caption{Same as Fig.~\ref{fig:p128bb_Sconst} but for the simulation Set~F.}
\label{fig:p128bb_Oconst}
\end{figure}

In Set~F, we keep $\Omega$ constant and vary the magnitude of the 
shear. In this case,
however, the Coriolis number no longer stays exactly constant because
the rms-velocity is affected by the rather strong shear for the
extreme values of $|q|$. This also affects the anisotropy of the
turbulence which is much greater than in Set~E, see
Fig.~\ref{fig:p128bb_Oconst}. The trends of $\tqxx$ and $\tqyy$ are
again opposite to each other as in Set~E, but here the variation as a
function of $q$ is significantly greater. Similarly as in Set~E, the
$\tqzz$-component is less affected, although a weakly incresing trend
is seen for small $q$ and a clearly decreasing trend is seen for
positive values of $q$. The off-diagonal component $\tqxy$ changes
sign at $q=0$ where also $\Sh$ changes sign. However, it is clear that
$\qxy$ is not linearly proportional to $q$ as might naively be expected if the
stress is fully due to turbulent viscosity. Thus it is important to
try to sepatate the nondiffusive and diffusive contributions (see below).

\begin{figure}[t]
\centering
\includegraphics[width=0.45\textwidth]{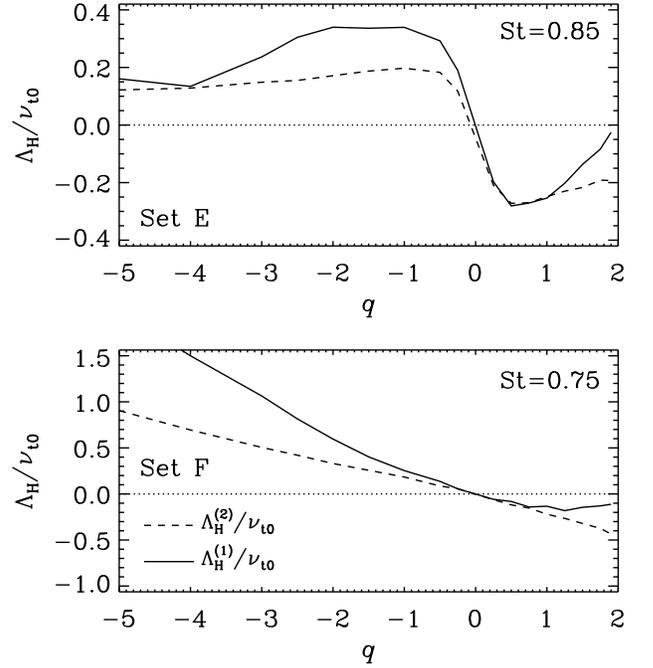}
\caption{Upper panel: the two measures of the horizontal
  $\Lambda$-effect from Set~E according to Eq.~(\ref{equ:L1}) (solid
  line) and Eq.~(\ref{equ:L2}) (dashed). Lower panel: the same as
  above but from Set~F.}
\label{fig:LambdaEF}
\end{figure}

\subsubsection{Closure model results}
\label{theta0results}

The Reynolds stresses obtained from the closure model are compared
with the results of the numerical simulations from Sets~E and F in
Figs.~\ref{fig:p128bb_Sconst} and
\ref{fig:p128bb_Oconst} with constant Sh and Co, respectively.
In both 
cases we observe rather good qualitative agreement between the numerical 
simulations and the closure model with the exception of $\tqzz$ in 
Set~F when $q<-2$ (see the third panel of Fig.~\ref{fig:p128bb_Oconst}). 
Another discrepancy
for $\tqzz$ occurs in the Set~E near $q=0$ where the simulation
results indicate a higher value than what is obtained from the closure
model. The discrepancies for $\tqzz$ occur for rapid rotation (Set~E
near $q=0$) and large shear (Set~F for $q<-2$) which suggests that the
simple closure model used here could be inadequate in such regimes.
We also note that the fit generally tends to get worse as $q$
approaches $2$. Partially due to this we cannot ascribe a single
Strouhal number that would fit the simulation data in the full
parameter range studied here. In Set~E the data is within error bars
consistent with $\St=0.75\ldots1.00$ in the range $q<0.5$ except for
$\qzz$ near $q=0$. Furthermore, $\St>1$ would be required to reproduce
$\qxx$ and $\qyy$ for $0.5<q<2$. In Set~F the situation is similar
with $\qzz$ and $\qxy$ deviating for large negative $q$ and $\qyy$
deviating for $q>1$.  All in all, the correspondence between the
simple closure and the simulation results is rather good.

\subsubsection{Separating diffusive and nondiffusive contributions}
\label{sec:seplanu}
When both, shear and rotation are present, the Reynolds stress can
also contain nondiffusive contributions proportional to the rotation
rate, i.e.\
\begin{equation}
Q_{ij} = \Lambda_{ijk}\Omega_k - \mathcal{N}_{ijkl} \mean{U}_{k,l},
\end{equation}
where $\Lambda_{ijk}$ describes the $\Lambda$-effect (Krause \&
R\"udiger \cite{KR74}; R\"udiger \cite{R80,R89}; Kitchatinov \& 
R\"udiger \cite{KR93}), which contributes
to the generation of differential rotation. In the case that we have
shear in addition to rotation, the problem is that for a given stress
component we have two unknown coefficients which makes it difficult to
distinguish the different contributions without a method
similar to the test field procedure in magnetohydrodynamics (see, e.g.\ 
Schrinner
et al.\ \cite{Shea05,Shea07}). However, if we consider that the
diffusive contribution is given by Eq.~(\ref{equ:nut}) with
$\nut=\St\,\nuto$, it is possible to separate the two provided that the
shear and rotation are sufficiently weak.

Considering the case $\bm\Omega=\Omega_0(0,0,1)$, the nondiffusive
part of the Reynolds stress can be described by a single coefficient
that is commonly denoted (cf.\ R\"udiger \cite{R89}) by
\begin{equation}
\qxy^{(\Omega)}=\Lambda_{\rm H} \cos\theta \Omega_0.\label{equ:LamH}
\end{equation}
This is often referred to as the horizontal $\Lambda$-effect because
it generates horizontal differential rotation in mean-field models of
stellar interior rotation. Let us now write the total stress as a sum
of the contributions from the $\Lambda$-effect and the turbulent
viscosity
\begin{equation}
\qxy = \lamh \Omega_0 - \nut S.
\end{equation}
The turbulent viscosity is given by $\nut=\onethird \tau \urms^2$ with
$\tau=\St/(\urms \kef)$, so $\nut=\onethird \St \urms \kef^{-1}=\St
\nuto$. 
Substituting this above we obtain
\begin{equation}
\qxy = \lamh \Omega_0 - \St \nuto S.
\end{equation}
Solving for $\Lambda_{\rm H}$ and dividing by $\nuto \Omega_0$ yields
\begin{equation}
\tlamh \equiv \tlamh^{(1)} = 6 \frac{\tqxy}{\Co}  - \St q,\label{equ:L1}
\end{equation}
where $\tilde{Q}_{xy}=\qxy/\urms^2$, $\tlamh = \lamh/\nuto$,
$q=-S/\Omega_0$ and where the definition of $\Co$,
Eq.~(\ref{equ:Corioliscomnew}) has been used.

On the other hand, for slow rotation and in the absence of shear the
nondiffusive stress due to the anisotropy of the turbulence can be written as
(R\"udiger \cite{R89}, see also KB08)
\begin{equation}
\qxy^{(\Omega)}=2\,\Omega_z \tau (\qyy-\qxx).
\end{equation}
Equating this with Eq.~(\ref{equ:LamH}) gives
\begin{equation}
\lamh=2\,\tau (\qyy-\qxx).
\end{equation}
Dividing by $\nuto$ and eliminating $\tau$ using $\St$ yields
\begin{equation}
\tlamh \equiv \tlamh^{(2)} = 6\,\St(\tqyy-\tqxx).\label{equ:L2}
\end{equation}
We can now compare the expressions~(\ref{equ:L1}) and (\ref{equ:L2})
by using the stresses and $\Co$ from the simulations and keeping the
Strouhal number as a free parameter. 

The results for the
two measures of the $\Lambda$-effect for the Sets~E and F are shown in
Fig.~\ref{fig:LambdaEF}. In Set~E the qualitative behaviour of the two
expressions is the same for Strouhal
numbers greater than $\approx0.7$, whereas the best correspondence 
between
$\tlamh^{(1)}$ and $\tlamh^{(2)}$ is obtained for
$\St\approx0.85$. For larger $\St$ the qualitative trend of
$\tlamh^{(1)}$ stays the same but the deviation between the two
expressions becomes increasingly greater for large values of $|q|$.
In summary, in Set~E we seem to be able to extract a reasonable
estimate of the nondiffusive contribution to the Reynolds stress with
the method outined above. The magnitude of $\lamh$ is of the order of
$(0.2\ldots0.4)\nuto$ when a similar Strouhal number is used in the fitting
as was required for the closure model to approximately reproduce the
simulation results. The magnitude of $\lamh$ is also quite close to 
the values obtained by
KB08 in a system without shear. 

In Set~F the qualitative behaviours of $\tlamh^{(1)}$ and
$\tlamh^{(2)}$ are the same for Strouhal numbers greater than about
$\approx0.5$. We find that in the Set~F a $q$-independent Strouhal
number does not give a very good fit to the numerical data. We have
plotted the curves for $\St=0.75$ which yields a reasonable fit to the
data in the vicinity of $q=0$ where $|\Sh|$ is small, and
Eqs.~(\ref{equ:L1}) and (\ref{equ:L2}) can be considered to be the least
affected by the shear.
Thus the results for the $\Lambda$-effect in Set~F leave more room for
speculation. One contributing factor is likely to be the significantly
stronger shear for the extreme values of $|q|$ which possibly renders
the expressions~(\ref{equ:L1}) and (\ref{equ:L2}) inaccurate in this
regime.

\onltab{3}{
   \begin{table*}
   \centering
   \caption[]{Summary of the runs in the Sets~G and H with $\theta=\pi/2$.}
   \vspace{-0.5cm}
      \label{tab:setGH}
     $$
         \begin{array}{p{0.03\linewidth}ccrrrccccrrr}
           \hline
           \noalign{\smallskip}
           Run      & $Grid$ & {\rm Re}  & q & \Co & \Sh & \urms & \tilde{Q}_{xx} & \tilde{Q}_{yy} & \tilde{Q}_{zz} & \tilde{Q}_{xy} & \tilde{Q}_{xz} & \tilde{Q}_{yz} \\
           \noalign{\smallskip}
           \hline
           \noalign{\smallskip}
           G1 &  128^3 &  24 &-5.00 &-0.06 & -0.16 & 0.062 & 0.330 & 0.346 & 0.325 & 0.053 & -0.007 &(-0.004) \\ % S005const_th90/128c
           G2 &  128^3 &  24 &-4.00 &-0.08 & -0.16 & 0.062 & 0.330 & 0.346 & 0.325 & 0.053 & -0.010 &(-0.007) \\ % 128c1
           G3 &  128^3 &  24 &-3.00 &-0.11 & -0.16 & 0.062 & 0.331 & 0.343 & 0.327 & 0.049 & -0.011 & -0.004  \\ % 128c2
           G4 &  128^3 &  24 &-2.50 &-0.13 & -0.16 & 0.062 & 0.332 & 0.342 & 0.327 & 0.048 & -0.012 & -0.004  \\ % 128c3
           G5 &  128^3 &  24 &-2.00 &-0.16 & -0.16 & 0.062 & 0.329 & 0.343 & 0.327 & 0.045 & -0.012 &(-0.001) \\ % 128c4
           G6 &  128^3 &  24 &-1.50 &-0.21 & -0.16 & 0.062 & 0.330 & 0.343 & 0.327 & 0.044 & -0.013 &(-0.001) \\ % 128c5
           G7 &  128^3 &  24 &-1.00 &-0.32 & -0.16 & 0.061 & 0.333 & 0.339 & 0.328 & 0.039 & -0.017 &(-0.001) \\ % 128c6
           G8 &  128^3 &  24 &-0.50 &-0.64 & -0.16 & 0.061 & 0.336 & 0.335 & 0.330 & 0.027 & -0.020 &(-0.000) \\ % 128c7
           G9 &  128^3 &  24 &-0.25 &-1.28 & -0.16 & 0.062 & 0.342 & 0.332 & 0.326 & 0.015 & -0.017 & (0.002) \\ % 128c8
          G10 &  128^3 &  24 & 0.25 & 1.27 & -0.16 & 0.062 & 0.342 & 0.332 & 0.327 & 0.016 &  0.015 & (0.000) \\ % 128c9
          G11 &  128^3 &  24 & 0.50 & 0.64 & -0.16 & 0.062 & 0.332 & 0.337 & 0.331 & 0.029 &  0.015 &  0.003  \\ % 128c10
          G12 &  128^3 &  24 & 0.75 & 0.42 & -0.16 & 0.062 & 0.329 & 0.343 & 0.328 & 0.036 &  0.015 &  0.004  \\ % 128c11
          G13 &  128^3 &  24 & 1.00 & 0.32 & -0.16 & 0.062 & 0.331 & 0.341 & 0.328 & 0.040 &  0.013 &  0.005  \\ % 128c12
          G14 &  128^3 &  24 & 1.25 & 0.25 & -0.16 & 0.062 & 0.329 & 0.340 & 0.329 & 0.042 &  0.012 &  0.005 \\ % 128c13
          G15 &  128^3 &  25 & 1.50 & 0.21 & -0.16 & 0.062 & 0.330 & 0.342 & 0.329 & 0.044 &  0.011 &  0.004  \\ % 128c14
          G16 &  128^3 &  24 & 1.75 & 0.18 & -0.16 & 0.062 & 0.330 & 0.343 & 0.328 & 0.045 &  0.011 &  0.005 \\ % 128c15
          G17 &  128^3 &  24 & 1.90 & 0.17 & -0.16 & 0.062 & 0.330 & 0.345 & 0.326 & 0.046 &  0.009 &  0.004 \\ % 128c16
           \hline
           \noalign{\smallskip}
           H1 &  128^3 &  35 &-5.00 &  0.22 &  0.55 & 0.088 & 0.257 & 0.424 & 0.328 &-0.148 & -0.024 & (0.006) \\ % O005cst_th90/128cc
           H2 &  128^3 &  31 &-4.00 &  0.25 &  0.49 & 0.080 & 0.276 & 0.397 & 0.333 &-0.129 & -0.026 &  0.005  \\ % 128cc1
           H3 &  128^3 &  30 &-3.00 &  0.26 &  0.39 & 0.076 & 0.302 & 0.374 & 0.327 &-0.116 & -0.034 & (0.015) \\ % 128cc2
           H4 &  128^3 &  28 &-2.50 &  0.28 &  0.35 & 0.071 & 0.314 & 0.359 & 0.328 &-0.098 & -0.033 &  0.013  \\ % 128cc3
           H5 &  128^3 &  26 &-2.00 &  0.29 &  0.29 & 0.066 & 0.326 & 0.344 & 0.330 &-0.077 & -0.029 &  0.010  \\ % 128cc4
           H6 &  128^3 &  25 &-1.50 &  0.31 &  0.23 & 0.064 & 0.337 & 0.336 & 0.327 &-0.058 & -0.023 &  0.009  \\ % 128cc5
           H7 &  128^3 &  24 &-1.00 &  0.32 &  0.16 & 0.062 & 0.335 & 0.336 & 0.330 &-0.038 & -0.018 &  0.004  \\ % 128cc6
           H8 &  128^3 &  24 &-0.50 &  0.33 &  0.08 & 0.060 & 0.336 & 0.333 & 0.331 &-0.016 & -0.009 &  0.003  \\ % 128cc7
           H9 &  128^3 &  24 &-0.25 &  0.32 &  0.04 & 0.060 & 0.338 & 0.332 & 0.340 &-0.008 & -0.008 &  0.003  \\ % 128cc8
          H10 &  128^3 &  24 & 0.25 &  0.33 & -0.04 & 0.060 & 0.337 & 0.333 & 0.331 & 0.011 &(0.002) &  0.003  \\ % 128cc9
          H11 &  128^3 &  24 & 0.50 &  0.33 & -0.08 & 0.060 & 0.336 & 0.335 & 0.329 & 0.021 &  0.007 &  0.002  \\ % 128cc10
          H12 &  128^3 &  24 & 0.75 &  0.32 & -0.12 & 0.061 & 0.334 & 0.335 & 0.331 & 0.029 &  0.011 & (0.003) \\ % 128cc11
          H13 &  128^3 &  24 & 1.00 &  0.32 & -0.16 & 0.061 & 0.333 & 0.337 & 0.330 & 0.040 &  0.013 & (0.004) \\ % 128cc12
          H14 &  128^3 &  25 & 1.25 &  0.31 & -0.19 & 0.063 & 0.327 & 0.348 & 0.326 & 0.050 &  0.018 &  0.005  \\ % 128cc13
          F15 &  128^3 &  25 & 1.50 &  0.31 & -0.23 & 0.064 & 0.334 & 0.342 & 0.324 & 0.063 &  0.021 &  0.009  \\ % 128cc14
          H16 &  128^3 &  26 & 1.75 &  0.30 & -0.26 & 0.066 & 0.331 & 0.344 & 0.325 & 0.075 &  0.025 &  0.013  \\ % 128cc15
          H17 &  128^3 &  26 & 1.90 &  0.30 & -0.28 & 0.066 & 0.330 & 0.342 & 0.328 & 0.074 &  0.025 &  0.011  \\ % 128cc16
          \hline
         \end{array}
     $$ 
   \end{table*}
}

\begin{figure*}[t]
\centering
\includegraphics[width=0.95\textwidth]{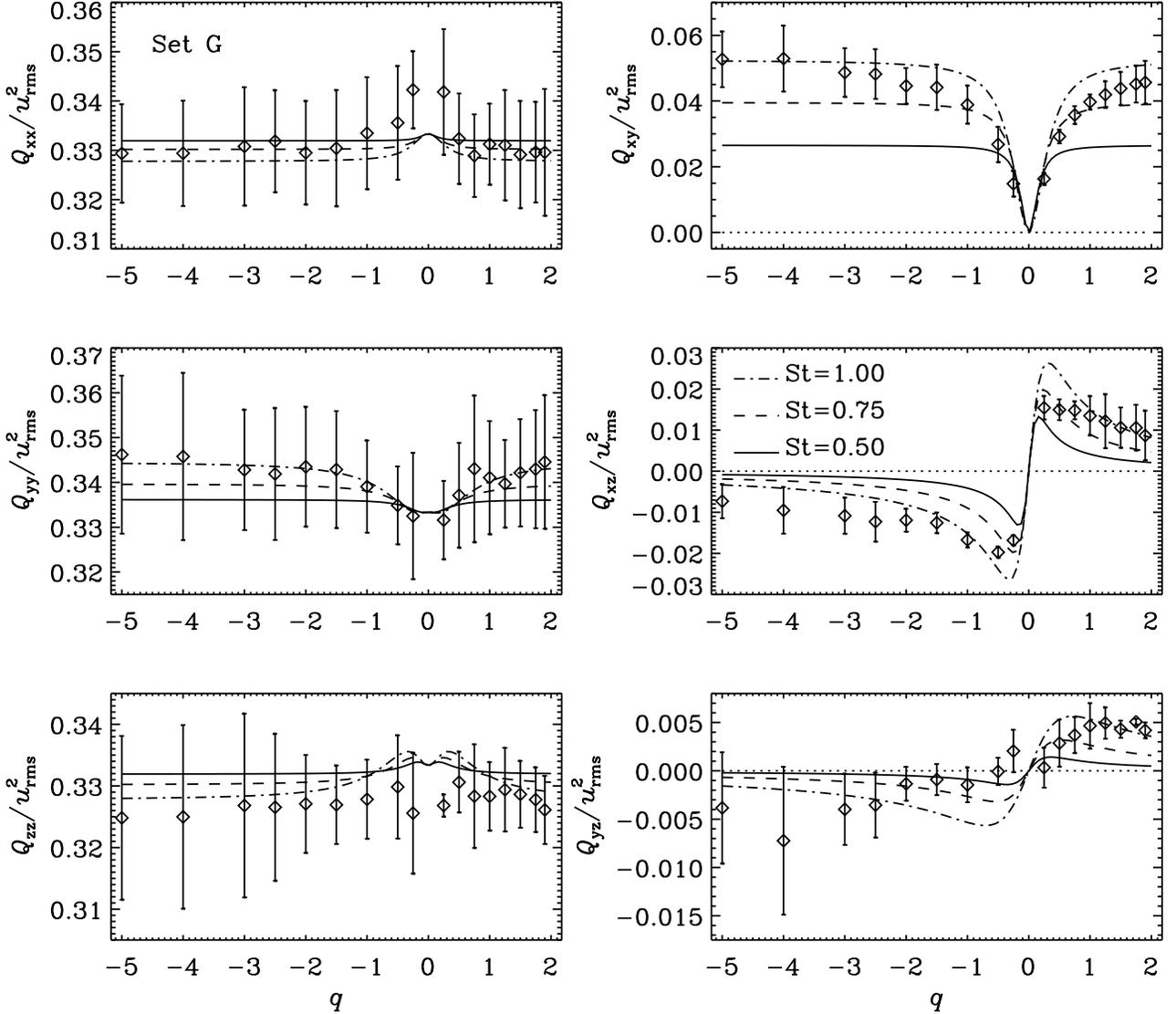}
\caption{Left column: $\tqxx$ (top panel), $\tqyy$ (middle), and
  $\tqzz$ (bottom) from the simulation Set~G. Right column: $\tqxy$
  (top panel), $\tqxz$ (middle), and $\tqyz$ (bottom) from Set~G. The
  curves in each panel show the closure model results with three
  different Strouhal numbers as indicated in the legend (righ column,
  middle panel).}
\label{fig:p128cc_Sconst}
\end{figure*}

\begin{figure*}[t]
\centering
\includegraphics[width=0.95\textwidth]{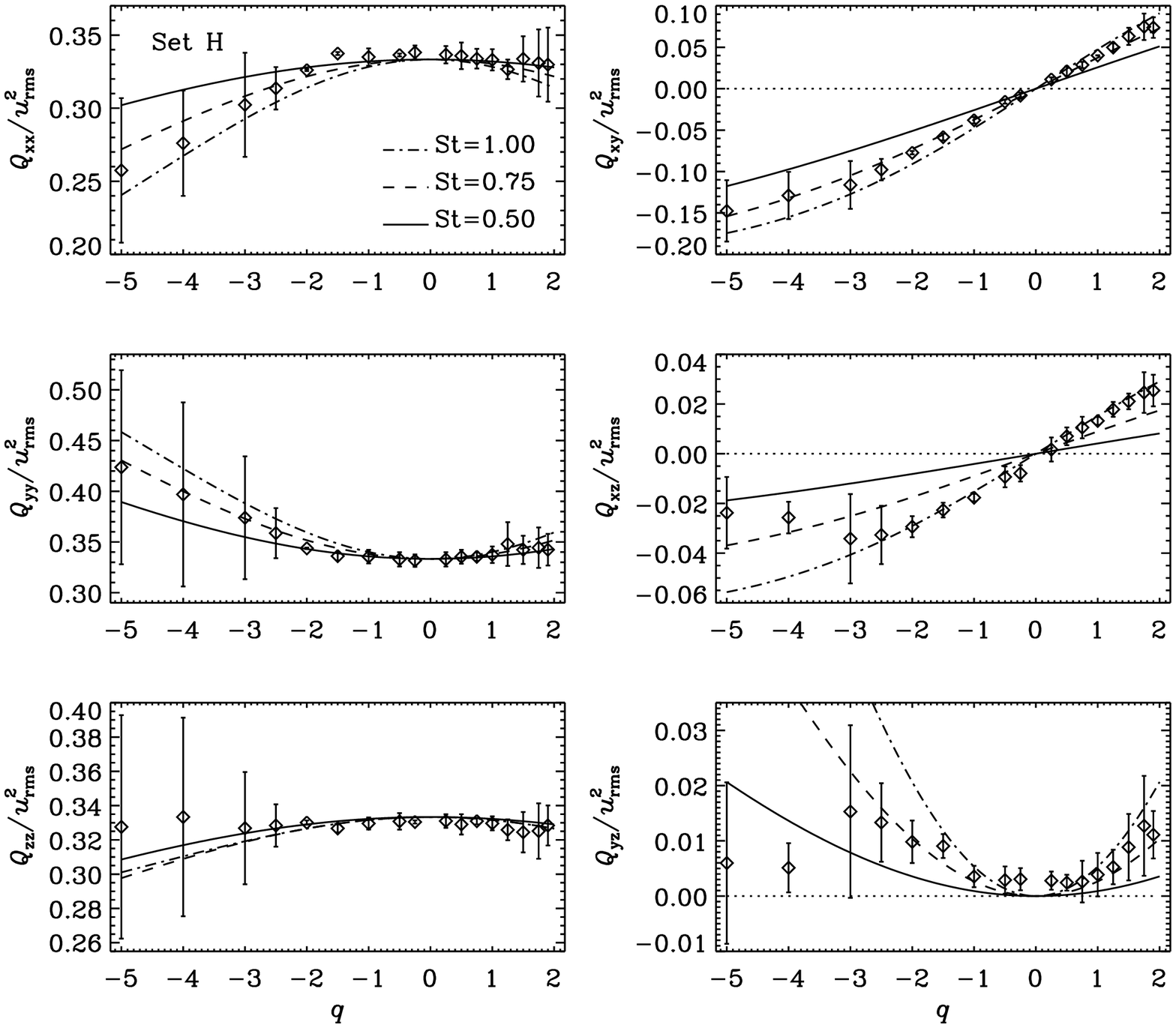}
\caption{Same as Fig.~\ref{fig:p128cc_Sconst} but for the simulation
  Set~H.}
\label{fig:p128cc_Oconst}
\end{figure*}

\subsection{Case $\Omega\neq0$, $\theta=90\degr$}
Finally, we consider the case where, in addition to the shear flow,
rotation with $\theta=90\degr$ is used, i.e.\
$\bm\Omega=\Omega_0(-1,0.0)$.
As is noted by Leprovost \& Kim (\cite{LK08b}) this system is unstable
even in the absence of turbulence. The instability generates
large-scale vorticity due to the fact that no stationary solution to 
the equations
of the large-scale velocity exist (Leprovost \& Kim \cite{LK08b}, see
also Appendix~\ref{sec:foba}) in the absence of forcing on the
large scales. If the system is interpreted as the polar region in the
solar tachocline, such large-scale forcing can be available, e.g.\, by
thermally driven flows but in the present simulation setup such
effects are difficult to implement.

In spite of the unstable nature of the system, we find that when
turbulence due to the forcing is present we can still extract
information about the Reynolds stresses from the early stages of the
simulations where the large-scale flows are still weak. The procedure
is similar to the case where rotation was absent (see
Sect.~\ref{sec:simnoom}). However, this means that the error bars in
Figs.~\ref{fig:p128cc_Sconst} and \ref{fig:p128cc_Oconst} are greater 
than those in
Figs.~\ref{fig:p128bb_Sconst} and \ref{fig:p128bb_Oconst} because much
shorter averaging intervals must be used.
The situation is aggravated in Set~H (see,
Fig.~\ref{fig:p128cc_Oconst}) where the amplitude of the shear flow
increases when $|q|$ increases.

\subsubsection{Simulation results}
We perform again two sets of simulations (see Table~\ref{tab:setGH})
where either the shear flow (Set~G) or rotation (Set~H) is kept
fixed. In the previous cases the source of anisotropy was either only
large-scale shear with mean vorticity
$\meanv{W}=\CURL{\meanv{U}}=S\hat{\bm z}$ (Sets~A to D), or shear and
rotation of the form $\bm\Omega=\Omega_0\hat{\bm z}$ (Sets~E and
F). In the present case the rotation vector points to the negative
$x$-direction due to which further symmetries are broken and the other
off-diagonal stresses $\qxz$ and $\qyz$ can have nonzero values (e.g.\
R\"udiger \cite{R89}).

The results from Set~G are shown in Fig.~\ref{fig:p128cc_Sconst}. Due
to the large error bars, the results for the diagonal components 
are consistent with a value
independent of $q$. The component $\qxx$ seems to increase and $\qyy$
to decrease near $q=0$, but these results are not statistically
significant. The off-diagonal stress $\qxy$ shows a similar
qualitative and quantitative trend as in Set~E. The component $\qxz$
is negative (positive) for $q<0$ ($q>0$) with absolute values peaking
near $q=0$. The magnitude of $\qxz$ is approximately one third of
$\qxy$. The $\qyz$ component is smaller by another factor of two to
three in comparison to $\qxz$ with a profile consistent with linear
proportionality to $q$. However, the large errors for $q\le0.25$ 
imply that the results there are not statistically significant.

The results from Set~H are shown in Fig.~\ref{fig:p128cc_Oconst}. The
diagonal stresses show again little dependence on $q$ except for
$q<-2.5$, where $\qxx$ decreases and $\qyy$ increases, although the
error bars are again so large that the results for $\qyy$ are also
consistent with a constant profile. Within error estimates,
$\qzz$ is consistent with a constant profile as a function of $q$. In
general, the results for $q<-2.5$ are rather unreliable due to the
rapid generation of large-scale flows. The off-diagonal stress $\qxy$
has a similar profile and somewhat larger magnitude than in
Set~F. Since the shear flow in both sets is the same the difference
can be due to the different nondiffusive contributions. Component
$\qxz$ shows a similar profile, and a magnitude of about one third of
the stress $\qxy$. $\qyz$, on the other hand, exhibits a profile
symmetric around $q=0$, apart from a few unrealiable points in the
regime $q<-2.5$.

\subsubsection{Closure model results}

The Reynolds stresses obtained from the closure model are compared
with the results from the numerical simulations of the Sets~G and H in
Figs.~\ref{fig:p128cc_Sconst} and \ref{fig:p128cc_Oconst},
respectively.
In Set~G the error bars for the diagonal components are so large that
a wide range of Strouhal numbers are consistent with the
results. Because the shear in this set is rather weak in all runs, 
the anisotropy of the turbulence
remains weak in all simulations. It also appers that the
simulation results for $\tqxx$ and $\tqzz$ are not captured by
the closure model for some points near $q=0$. The off-diagonal
components $\tqxy$ and $\tqxz$ show a clearer picture and the closure
model is consistent with the numerical results if
$\St=0.75\ldots1$. The simulation results for $\tqyz$ are for the most
part not distinguishable from zero for $q<0.5$. For $q$ greater than
this, the closure model is consistent with the simulations for
$\St=1$.

In Set~H the diagonal components show weak signs of anisotropy if
$|q|<2$. However, the errors increase substantially for greater $q$
and the data is still almost consistent with isotropic turbulence
even for $q=-5$. This is due to the short averaging interval that
results in from the vigorous vorticity generation that is excited very early in
these runs. This is because the extreme values of $|q|$ correspond to 
large values of $\Sh$ in this
set. All of the off-diagonal stresses show qualitatively correct behaviours
and the simulation data is again consistent with $\St=0.75\ldots1$ for
$|q|<3$.

\begin{figure}[t]
\centering
\includegraphics[width=0.45\textwidth]{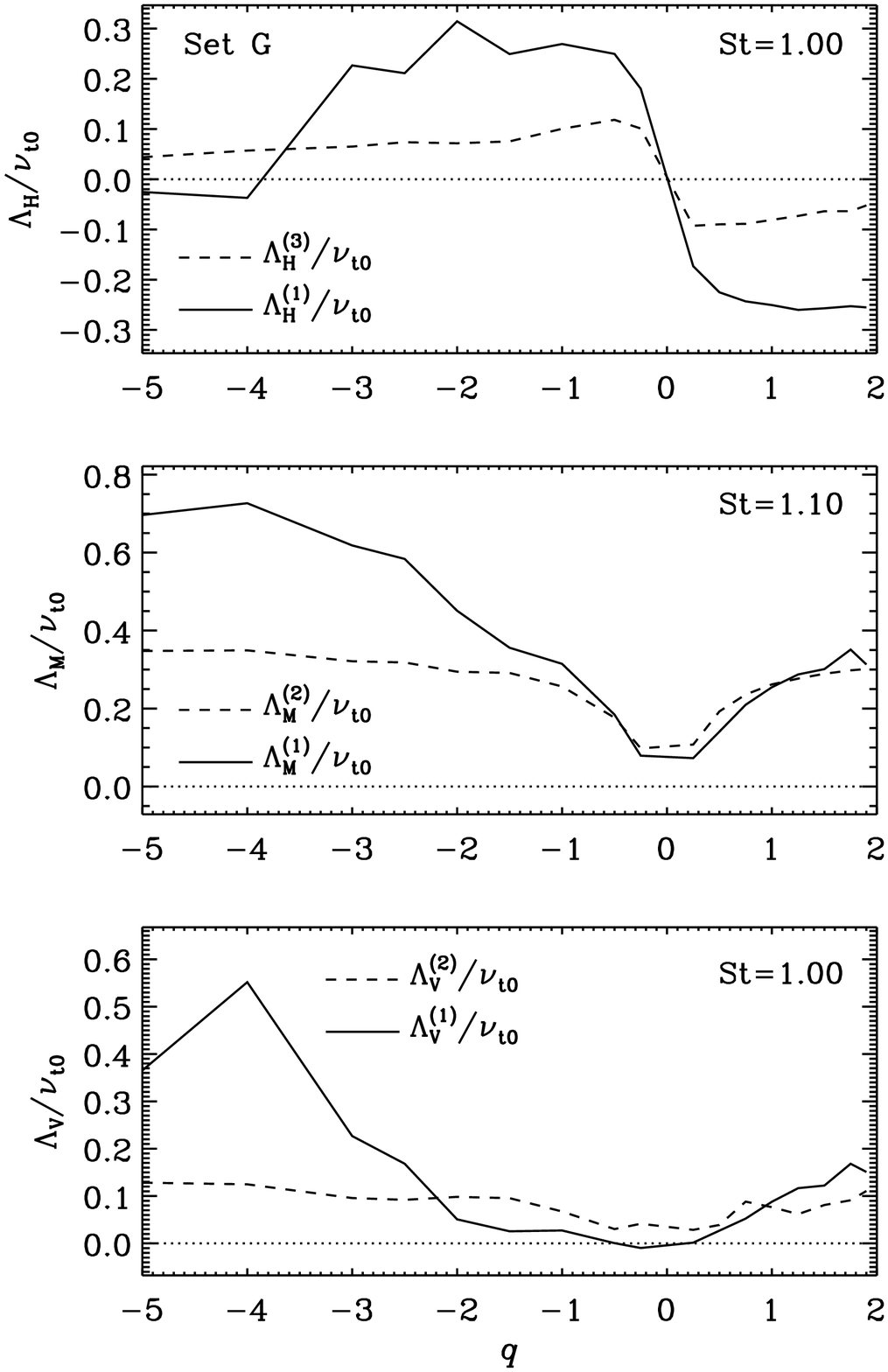}
\caption{Nondiffusive contributions to the Reynolds stresses
  parameterized by the coefficients $\tlamh$ (top panel), $\tlamm$
  (middle), and $\tlamv$ (bottom) from the simulation Set~G. In each
  panel two expressions are compared and the linestyles are explained
  in the legends.}
\label{fig:LambdaG}
\end{figure}

\begin{figure}[t]
\centering
\includegraphics[width=0.45\textwidth]{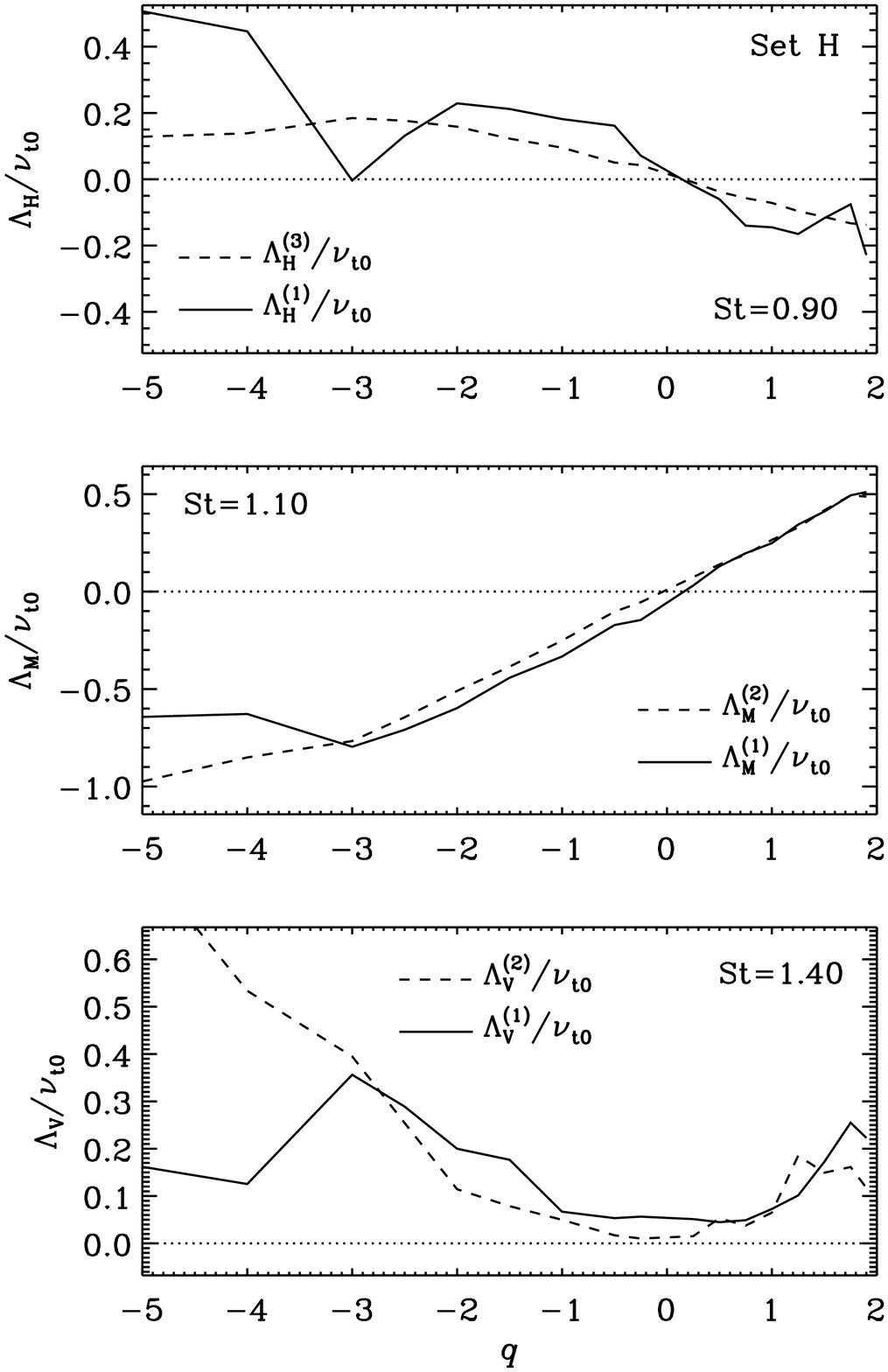}
\caption{Same as Fig.~\ref{fig:LambdaG} but for the simulation Set~H.}
\label{fig:LambdaH}
\end{figure}

\subsubsection{Separating diffusive and nondiffusive contributions}
In the case with $\theta=90\degr$, Eq.~(\ref{equ:LamH}) no longer
applies, but we can still use Eq.~(\ref{equ:L1}) to estimate the
nondiffusive contribution $\tlamh^{(1)}$. On the other hand
Eq.~(\ref{equ:L2}) is no longer valid because $\Omega_z=0$. To replace
Eq.~(\ref{equ:L2}), we can use the tau-approximation to derive another
expression for the nondiffusive stress. We arrive at a term
proportional to $\Omega_x$ in the equation of $\qxy$
\begin{equation}
\qxy^{(\Omega)}=2\,\Omega_x \tau \qxz \equiv \lamh^{(3)} \Omega_0,
\end{equation}
where the superscript differentiates this contribution from those used
in the case where $\theta=0$. Division by $\nuto$ yields
\begin{equation}
\tlamh^{(3)} = - 6\, \St \tqxz.\label{equ:L3}
\end{equation}
Furthermore, to lowest order, the other two off-diagonal stresses can
contain nondiffusive contributions
\begin{eqnarray}
\qxz^{(\Omega)} &=& \lamm \Omega_0, \label{equ:M1}\\
\qyz^{(\Omega)} &=& \lamv \sin \theta \Omega_0.\label{equ:V1}
\end{eqnarray}
Solving for the $\Lambda$-coefficients and normalizing by $\nuto$
gives
\begin{eqnarray}
\tlamm &\equiv& \tlamm^{(1)} = -6 \frac{\tqxz}{\Co}, \\
\tlamv &\equiv& \tlamv^{(1)} = -6 \frac{\tqyz}{\Co}.\label{equ:tlamv1}
\end{eqnarray}
We note that whilst Eq.~(\ref{equ:V1}) coincides with that commonly
assumed for the $\Lambda$-effect, Eq.~(\ref{equ:M1}) does
not. Typically the ``meridional'' $\Lambda$-effect, $\Lambda_{\rm M}$,
is proportional to $\sin\theta \cos\theta \Omega$ (e.g.\ Pulkkinen et
al.\ \cite{Pea93}; K\"apyl\"a et al.\ \cite{KKT04}) which
derives from symmetry considerations in a rotating spherical
system. However, here these symmetry considerations are no longer 
obeyed due to the
added shear flow.
Furthermore, applying the minimal tau-approximation to the Reynolds
stress equation yields
\begin{eqnarray}
\qxz^{(\Omega)} &=&-2\,\Omega_x \tau \qxy, \\
\qyz^{(\Omega)} &=& 2\,\Omega_x \tau (\qzz-\qyy).
\end{eqnarray}
Equating these with Eqs.~(\ref{equ:M1}), and
(\ref{equ:V1}), respectively, and bearing in mind that
$\Omega_x=-\Omega_0$, we obtain, after normalization with the turbulent
viscosity $\nuto$,
\begin{eqnarray}
\tlamm^{(2)} &=&  6\,\St\tqxy,\\
\tlamv^{(2)} &=& -6\,\St(\tqzz-\tqyy).
\end{eqnarray}
We now proceed with the comparison in the same manner as in 
Sect.~\ref{sec:seplanu}.

We find that in Set~G the correspondence between the two measures for
each of the $\Lambda$-coefficients are in rough qualitative agreement
for $\St\approx1$, but matching the quantitative values is not possible
(see Fig.~\ref{fig:LambdaG}). Compared to Set~E, the profile and
amplitude of $\tlamh\approx0.05\ldots0.3$ are similar, although a
precise value is difficult to determine due to the disagreement between
the two expressions. The other two nondiffusive components have
magnitudes similar to $\tlamh$ at least in the range $|q|<2$. 
For $\tlamm$ the analytical expressions seem to agree rather well in
the range $|q|<2$. For $\tlamv$, on the other hand, the correspondence
is not very good, although the sign of the coefficient is for the most
part reproduced correctly.
Here, however, the indeterminate nature of $\tqyz$ is likely to affect
the results from Eq.~(\ref{equ:tlamv1}).

In Set~H the correspondence between the different expressions for the 
$\Lambda$-coefficients is somewhat better. For $q<-2$ the
simulations begin to exhibit vigorous vorticity generation at a
very early stage making the error estimates increase significantly and
the results for the $\Lambda$-effect in this regime are not very
reliable. Again the best fit is achieved with $\St\approx1$, although
a unique value that would fit all components cannot be found. The
maxima of all $\Lambda$-coefficients are of the order 0.5 in this set.

\subsection{On the validity of the MTA} 
\begin{figure}[t]
\centering
\includegraphics[width=0.45\textwidth]{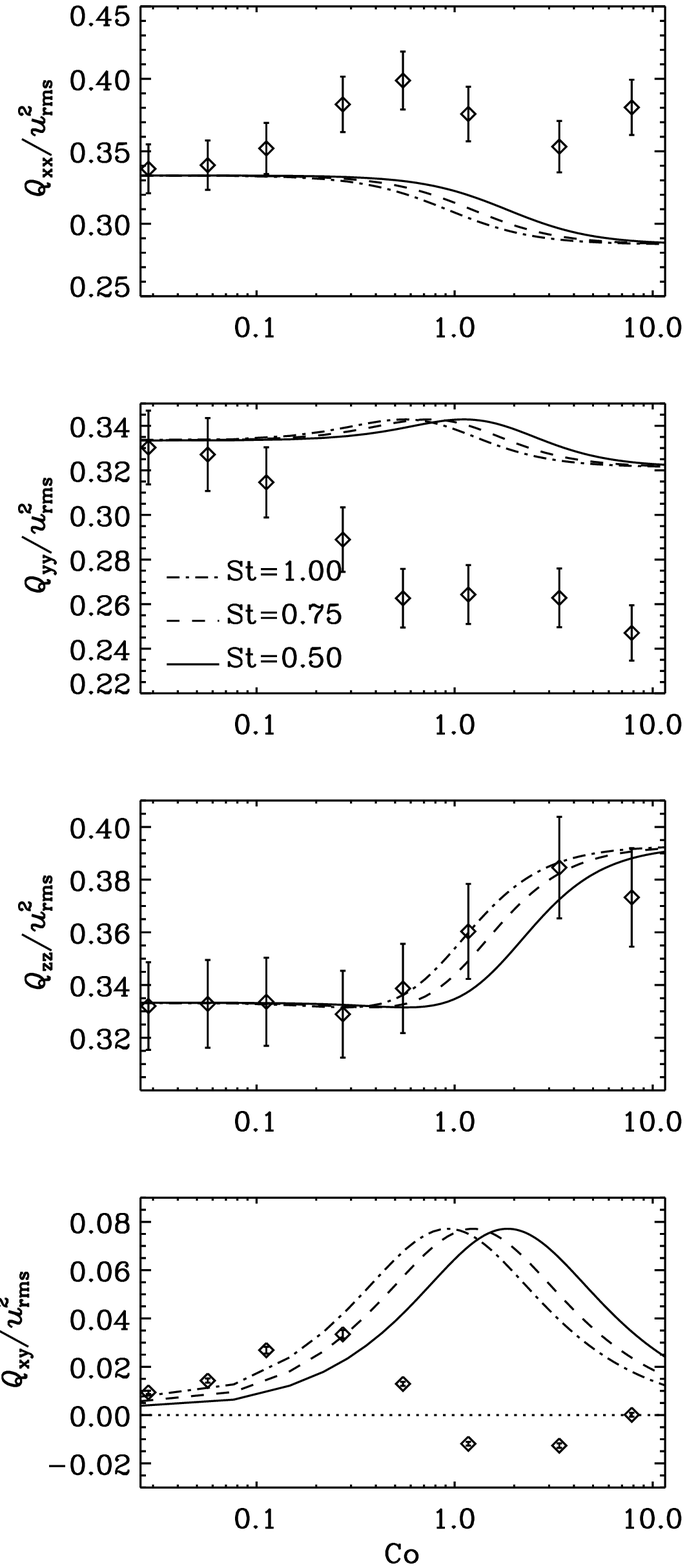}
\caption{ Same as Fig.~\ref{fig:p128bb_Sconst} but for a set of
  runs with $q=1=\mbox{const}$ and $\Rey\approx28$.}
\label{fig:pqconst}
\end{figure}

 We note that the largest deviations between the simulations and
  the closure model are seen when rapid rotation or strong shear are
  used. This may suggest that the simple closure model used here can
  break down at such circumstances.
 In order to test the validity of the present formulation of the MTA, 
we perform a set of
  simulations where $q=1$ is constant and the values of $\Co$ and
  $\Sh$ are varied. These results are compared to those from the
  MTA-closure model where the relaxation time is independent of $\Co$
  and $\Sh$.
  The results are shown in Fig.~\ref{fig:pqconst}. It appears that the
  simulation results for all of the stress components are in
  accordance with the closure model only for $\Co\la0.1$. The components $\qxx$
  and $\qyy$ are not well reproduced for greater
  $\Co$. Rather surprisingly the $\qzz$ component is consistent with the
  closure model for all values of $\Co$ explored here. For $\qxy$
  the agreement is rather good for $\Co\la0.3$, but for greater $\Co$ the
  stress changes sign as a function of $\Co$ in the simulations which
  is not reproduced by the closure model. Due to the differing
  behaviour of the stresses we cannot estimate the validity range of
  the MTA-closure very precisely based on this data only.
It is likely that with strong enough rotation or shear the 
relevant time scale to be used in the closure model is associated with 
rotation or shear instead of our naive 
estimate of the eddy turnover time Eq.~(\ref{equ:St}). For
  comparison, the time scales related to rotation and shear are of the
  same order of magnitude as the turnover time of the turbulence when
  $\Co=0.5$. At face value, the present results suggest that the
  effects of rotation and shear begin to affect the results even at
  somewhat slower rotation.

\begin{figure}[t]
\centering
\includegraphics[width=0.45\textwidth]{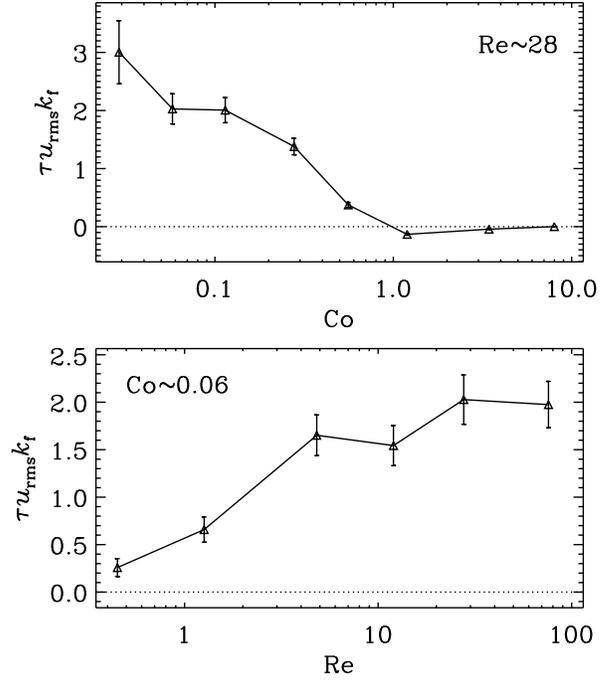}
\caption{ Upper panel: Strouhal number from the triple correlations
  as a function of $\Co$ for the same runs as in
  Fig.~(\ref{fig:pqconst}). Lower panel: $\St$ as a function of $\Rey$
  for runs with $\Co\approx0.06$ and $q=1$.}
\label{fig:ptau}
\end{figure}

  Furthermore, it turns out that in the presence of a shear flow of
  the form $\mean{U}_y(x)$ it is possible to estimate the relaxation
  time using the basic assumption of the MTA, Eq.~(\ref{equ:baMTA}),
  from the equation of $\qxy$. This is analogous to the passive
  scalar case where the flux due to an imposed gradient was found to
  the proportional to the triple correlation (Brandenburg et al.\
  \cite{BKM04}). Let us now write
\begin{equation}
\tau=-\frac{\qxy}{T_{xy}},
\end{equation}
where
\begin{equation}
T_{xy}=-c_s^2 \overline{ u_i \partial_j \ln \rho + u_j \partial_i \ln \rho} + \mean{\qij\DIV{\bm{u}}}.\label{equ:Txy}
\end{equation}
We find that the last term of Eq.~(\ref{equ:Txy}) is negligible in
comparison to the other terms in all cases considered here.
Representative results for the Strouhal number defined by
Eq.~(\ref{equ:St}) from two sets of runs are shown in
Fig.~\ref{fig:ptau}. Firstly, as a function of rotation, the Strouhal
number is between 1.5 and 3 for $\Co\la0.3$, whereas the value
decreases for more rapid rotation and shear. The relaxation time can
even be negative for $\Co\ga1$, indicating that the basic assumption
of the MTA breaks down. Furthermore, for weak rotation and shear, the
Strouhal number is constant as a function of the Reynolds number
provided that $\Rey>1$.
This is consistent with the fact that in the regime $\Rey\approx1$ the
viscous time scale is of the same order of magnitude as the turnover
time and neglecting the viscous terms in the equation of the stress is
not justified. However, as we use Reynolds numbers of the order of
$30$ in most of our simulations, omitting the viscosity is permitted.

%______________________________________________________________

\section{Conclusions}
\label{sec:conclusions}
We have performed simulations of isotropically forced homogeneous
turbulence under the influences of shear and rotation in order to
study the turbulent transport properties. We find that in the absence
of rotation the turbulent viscosity is of the order of the first order
smoothing estimate $\nuto=\onethird\urms\kef^{-1}$, and that the
Strouhal number saturates to a value of $\St\approx1.5$ for large
Reynolds numbers.

When both rotation and shear are present, we seek to distinguish the
diffusive (turbulent viscosity) and nondiffusive parts
($\Lambda$-effect) of the stress. In the case of $\qxy$ this is
achieved by assuming the turbulent viscosity to be proportional to
$\St \nuto$ and solving for the $\Lambda$-part from the equation of
total stress, whereas for the other two off-diagonal components no
diffusive contributions are assumed to appear in the equation of 
the total stress. As an
independent check we derive analytical expressions for the
$\Lambda$-effect using the minimal tau-approximation. The two
expressions are then compared by using the Strouhal number as a free
parameter. Although the results for the $\Lambda$-effect from the
different approaches are not fully consistent in all cases, we find
that they agree at least qualitatively, if not quantitatively, in most
cases. The magnitude of the $\Lambda$-effect is a few times $0.1\nuto$ in
most cases in accordance with earlier studies (e.g. K\"apyl\"a \&
Brandenburg \cite{KB07,KB08}). However, a more precise determination
and separation of the different coefficients requires a method akin to
the test field method used succesfully in magnetohydrodynamics
(Schrinner et al.\ \cite{Shea05,Shea07}).

We have also studied the Reynolds stresses using the minimal
tau-approximation closure model where the Strouhal number $\St$ is the
only free parameter. Comparing the results of this very simple closure
model with our simulations shows that they generally agree
surprisingly well with rather few exceptions. We find that in most
cases the best fit between the closure model and full simulations is
achieved when $\St \approx 1$. 
 We note, however, that the agreement between the closure model
  and the simulations becomes worse as the rotation rotation rate or
  shear are increased. The reason is likely to be that the relevant
  time scale in those regimes is no longer the turnover time but
  rather the time scale related to the rotation or shear. We feel that
  investigating the validity of the tau-approximation more precisely
  and futher improving its performance in reproducing the simulation
  results is a valid subject for further study.

Finding a closure model that reproduces the relevant properties of
turbulence would be very useful in a variety of astrophysical
applications. From the viewpoint of the present paper the most
interesting application is the angular momentum balance of
convectively unstable stellar interiors where rotation and shear flows
are known to play important roles. Although the present model is
highly idealised in comparison to stratified convection, the present
results using the MTA-closure are promising. A logical step towards
more general description of convective turbulent transport would be to
study Boussinesq convection (e.g.\ Spiegel \& Veronis \cite{SV60}) and
extend the closure model to work in the same regime (Miller \& Garaud
\cite{MG07}). In addition to the Reynolds stresses, such models could
be used to study anisotropic turbulent heat transport (e.g.\
R\"udiger \cite{R82};
Kitchatinov et al.\ \cite{KPR94}; Kleeorin \& Rogachevskii
\cite{KR06}). Furthermore, these models could help in answering how
some enigmatic results of local convection simulations,
such as the narrow regions of very strong stresses near the equator
(Chan \cite{C01}; K\"apyl\"a et al.\ \cite{KKT04}; R\"udiger et al.\
\cite{REZ05}) in the rapid rotation regime, can be explained. These 
subjects will be studied in
subsequent publications.

\begin{acknowledgements}
  We acknowledge the helpful comments from an anonymous referee. 
  The computations were performed on the facilities hosted by the
  CSC -- IT Center for Science in Espoo, Finland, who are financed by
  the Finnish ministry of education. The authors acknowledge financial
  support from the Academy of Finland grant Nos.\ 121431 (PJK) and
  112020 (JES, MJK, AJL). PJK and MJK acknowledge the hospitality of
  Nordita during the program `Turbulence and Dynamos.' JES 
  acknowledges the financial support from the Finnish Cultural Foundation.
\end{acknowledgements}

\appendix

\section{Stationary solutions to the MTA-closure model}
\label{sec:appendix}
Stationary solution to Eq.~(\ref{finalcl}) can be obtained by
setting $\dot{Q}_{ij}=0$ and solving the resulting equations
\begin{eqnarray}
\qxx &=& 4\,\Omega_z\tau Q_{xy}+ \qxx^{(0)}, \label{equ:1} \\
\qxy &=& -S \tau Q_{xx}+2\,\Omega_x \tau Q_{xz}+2\,\Omega_z \tau (Q_{yy}-Q_{xx}), \\
\qxz &=& 2\,\Omega_z \tau Q_{yz}-2\,\Omega_x \tau Q_{xy},\\
\qyy &=& -2\,S\tau Q_{xy}+4\,\Omega_x\tau Q_{yz}-4\,\Omega_z\tau Q_{xy} + \qyy^{(0)}, \\
\qyz &=& -(S+2\,\Omega_z)\tau Q_{xz} +2\,\Omega_x \tau (Q_{zz}-Q_{yy}),\\
\qzz &=& -4\,\Omega_x\tau Q_{yz}+ \qzz^{(0)}, \label{equ:6}
\end{eqnarray}
where $\Omega_x$ and $\Omega_z$ are the components of the rotation
vector as defined in Eq.~(\ref{OmegaDef}), and we have used the fact
that $Q_{xy}^{(0)} = Q_{xz}^{(0)} = Q_{yz}^{(0)} =0$.

\subsection{Only shear, $\Sh\neq 0$, $\Co=0$.}
\label{sucsec:appos}
In the absence of rotation, the stationary solution to
Eqs.~(\ref{equ:1}) to (\ref{equ:6}) is
\begin{eqnarray}
\qxx &=& \qxx^{(0)},\\
\qxy &=& -\St \Sh \qxx^{(0)},\\
\qyy &=& 2\,\St^2 \Sh^2 \qxx^{(0)} + \qyy^{(0)},\\
\qzz &=& \qzz^{(0)},\\
\qxz &=& \qyz = 0,\\
Q &=& \qxx^{(0)} (1+2\,\St^2 \Sh^2) + \qyy^{(0)} + \qzz^{(0)}.
\end{eqnarray}
Since the forcing in the present case is isotropic, we can write
\begin{equation}
\qxx^{(0)} = \qyy^{(0)} = \qzz^{(0)} = \onethird Q^{(0)},\label{equ:isof}
\end{equation}
where $Q^{(0)}$ is the trace of $\qij^{(0)}$ in the case without
rotation or shear. Substituting (\ref{equ:isof}) into the expression
of $Q$, we obtain
\begin{equation}
\urms^2 = Q = Q^{(0)} \left(1 + \twothirds \St^2 \Sh^2 \right).\label{equ:urms2}
\end{equation}
Using Eq.~(\ref{equ:isof}) and (\ref{equ:urms2}) the normalised
stresses, $\tilde{Q}_{ij}=\qij/\urms^2$ can be written in terms of the
dimensionless parameters $\St$ and $\Sh$ as
\begin{eqnarray}
\tqxx &=& \tqzz = \frac{1}{3+2\,\St^2 \Sh^2},\\
\tqxy &=& \frac{-\St \Sh}{3+2\,\St^2 \Sh^2},\\
\tqyy &=& 1 - \frac{2}{3+2\,\St^2 \Sh^2}.
\end{eqnarray}

\subsection{Shear and rotation, $\Sh, \Co\neq0$, $\theta=0$.}

If rotation is present in the system and $\theta=0$, the solution of
Eqs.~(\ref{equ:1}) to (\ref{equ:6}) can be written as
\begin{eqnarray}
Q_{xx} &=& \frac{[2\, \Co \St^2 (\Co + \Sh)+1] Q_{xx}^{(0)}+2\, \Co^2 \St^2 Q_{yy}^{(0)}}{1+4\, \Co \St^2 (\Co+\Sh)}, \\
Q_{xy} &=& \frac{-(\Co + \Sh) \St Q_{xx}^{(0)}+\Co \St Q_{yy}^{(0)}}{1+4\, \Co \St^2 (\Co+\Sh)}, \\
Q_{yy} &=& \frac{[2\, \Co \St^2 (\Co + \Sh)+1] Q_{yy}^{(0)}}{1+4\, \Co \St^2 (\Co+\Sh)} \nonumber \\ 
 & & + \frac{2\, (\Co+\Sh)^{2} Q_{xx}^{(0)}}{1+4\, \Co \St^2 (\Co+\Sh)}, \\
Q_{zz} &=& Q_{zz}^{(0)}, \\
Q_{xz} &=& Q_{yz}=0,\\
Q &=& \left(1 + \frac{2\, \Sh^2 \St^2}{3 + 12\Co \St^2 (\Co+\Sh)}\right)Q^{(0)}.
\end{eqnarray}
If we allow $\tau \neq\tau_{\rm f}$ the solution can be obtained by
simply multiplying the above solutions with a factor $\tau/\tau_{\rm
  f}$.

\subsection{Shear and rotation, $\Sh, \Co\neq0$, $\theta=\pi/2$.}

Considering the case $\theta=\pi/2$, the solution is
\begin{eqnarray}
Q_{xx} & = & Q_{xx}^{(0)}, \\
Q_{xy} & = & -\frac{Q_{xx}^{(0)} \Sh \St}{1+ \Co^2 \St^2}, \\
Q_{xz} & = & -\frac{   Q_{xx}^{(0)} \St^2 \Co \Sh}{1+ \Co^2 \St^2}, \\
Q_{yy} & = & \frac{ 2\, \Sh^2 \St^2 Q_{xx}^{(0)}\!+\!Q_{yy}^{(0)}+\!2\,( Q_{yy}^{(0)} + Q_{zz}^{(0)} )\Co^2 \St^2}{1+4\, \Co^2 \St^2}, \\
Q_{yz} & = & \frac{ [(Q_{yy}^{(0)} - Q_{zz}^{(0)}) \Co^2 \St^2 ] \Co \St}{(1+4\, \Co^2 \St^2)(1+ \Co^2 \St^2)} \nonumber \\ 
& & \hspace{1cm} + \frac{(3 \Sh^2 Q_{xx}^{(0)} \St^2-Q_{zz}^{(0)}+Q_{yy}^{(0)}) \Co \St}{(1+4\, \Co^2 \St^2)(1+ \Co^2 \St^2)}, \\
Q_{zz} & = & \frac{ 2\, (Q_{yy}^{(0)} + Q_{zz}^{(0)}) \Co^4 \St^4 + 6 \St^4 Q_{xx}^{(0)} \Co^2 \Sh^2 }{(1+4\, \Co^2 \St^2)(1+\Co^2 \St^2)} \nonumber \\ 
 & & + \frac{3 Q_{zz}^{(0)}\Co^2 \St^2+2\, Q_{yy}^{(0)} \Co^2 \St^2+Q_{zz}^{(0)}}{(1+4\, \Co^2 \St^2)(1+\Co^2 \St^2)}, \\
Q &=& \left(1+\frac{2}{3}\frac{ \Sh^2 \St^2}{1+\Co^2 \St^2}\right)Q^{(0)}.
\end{eqnarray}
If $\tau \neq\tau_f$, the solution can be obtained in the same way as
in the $\theta=0$ case.

\section{Stability analysis}
\subsection{Vorticity dynamo in the presence of rotation}
\label{sec:vortdyn}
The existence of the vorticity dynamo in the present case can be
studied using the same procedure used in K\"apyl\"a et al.\
(\cite{KMB09}). When one assumes that the average velocity field
depends only on $z$, the relevant mean field equation has the form
\begin{equation}
\dot{\meanv{U}} = -S \mean{U}_x \hat{\bm y} - \overline{\bm{u} \cdot \bm\nabla\bm{u}} - 2\,\bm\Omega \times \meanv{U} + \nu {\meanv{U}}'',
\label{equ:mvrd}
\end{equation}
where the primes denote $z$-derivatives. Assuming the flow to be
incompressible, $\bm\nabla \cdot \meanv{U}=\meanv{U}_{z,z}=0$, so
$\mean{U}_z=\mbox{const}=0$ by a suitable choice of the initial
condition. The nonlinear term $-\mean{\bm{u} \cdot \bm\nabla \bm{u}}$
can be represented in terms of the large-scale velocity by introducing
a turbulent eddy viscosity tensor $\nu_{ij}$ (see, e.g.\ Elperin et
al.\ \cite{EKR03}; K\"apyl\"a et al.\ \cite{KMB09})
\begin{equation}
-(\mean{\bm{u} \cdot \bm\nabla \bm{u}})_i = \nu_{ij} \mean{U}''_j.\label{equ:stress}
\end{equation}
Substituting Eq.~(\ref{equ:stress}) to Eq.~(\ref{equ:mvrd}) gives 
\begin{eqnarray}
\dot{\mean{U}}_x &=& (\nu+\nu_{xx}) {\mean{U}_x}''+\nu_{xy} {\mean{U}_y}''+ 2\,\Omega_z \mean{U}_y,\label{equ:cmvrd0}\\
\dot{\mean{U}}_y &=& (\nu+\nu_{yy}) {\mean{U}_y}''+\nu_{yx} {\mean{U}_x}''-(S+2\,\Omega_z) \mean{U}_x.
\label{equ:cmvrd}
\end{eqnarray}
We note that because $\mean{U}_z=0$, no terms proportional to $\Omx$
appear.
Similarly as in K\"apyl\"a et al.\ \cite{KMB09}, we assume a
stationary state and seek solutions to Eqs.~(\ref{equ:cmvrd0}) and
(\ref{equ:cmvrd}). Now we can apply a Fourier transform and arrive at
\begin{eqnarray}
(\nu+\nu_{xx})k^2 {\hat{U}_x}+(\nu_{xy}k^2-2\,\Omega_z) \hat{U}_y &=& 0, \label{equ:fmvrd0} \\
(\nu+\nu_{yy}) k^2 {\hat{U}_y}+(\nu_{yx}k^2+S+2\,\Omega_z) \hat{U}_x &=& 0,
\label{equ:fmvrd}
\end{eqnarray}
where $\hat{U}_i$ are the Fourier amplitudes of $\mean{U}_i$, and
$k$ is the wavenumber.
Equations~(\ref{equ:fmvrd0}) and (\ref{equ:fmvrd}) imply that the
sufficient condition for the mean vorticity dynamo to appear is
\begin{equation}
\frac{(\nu_{xy} - 2 \frac{\Omega_z}{k^2})(\nu_{yx}+\frac{S+2\Omega_z}{k^2})+\epsilon^2}{\nu_T^2} \ge 1,
\end{equation}
where $\nu_T=\nu+\frac{1}{2}(\nu_{xx}+\nu_{yy})$ and 
$\epsilon=\frac{1}{2}(\nu_{xx}-\nu_{yy})$. 
In the absence of shear we recover the expression derived in
K\"apyl\"a et al.\ \cite{KMB09}, suggesting that $\nu_{xy}S>0$.
On the other hand, if we set $\nu=0$ and $\nu_{ij}=0$, we have
\begin{equation}
2 \frac{\Omega_z^2}{k^4} (q-2) \ge 0,
\end{equation}
where $q=-S/\Omega_z$. Since $\Omega_z^2/k^4$ is always 
positive, this inequality holds if $q \ge 2$. This is the standard 
Rayleigh stability condition (for instability).
It is clear that rotation has a stabilizing effect in the system
provided that $q<2$. It is, however, conceivable that for small enough
$q$ the component $\nu_{xy}$ overcomes the rotational
stabilization. However, even for our most extreme runs with $q=-5$
this appears not to be the case.

\subsection{Force balance for $\Sh, \Co\neq0$, $\theta=\pi/2$}
\label{sec:foba}
We find that large-scale flows tend to always appear in the simulations at
$\theta=\pi/2$. A very similar setup was studied by Leprovost \& Kim
(\cite{LK08b}) who studied the force balance for the large-scale flow 
including also the pressure gradient
\begin{equation}
\dot{\meanv{U}} = -S \mean{U}_x \hat{\bm y} - \overline{\bm{u} \cdot \bm\nabla\bm{u}} -2\,\bm\Omega \times \meanv{U}-\frac{1}{\rho}\bm\nabla P + \nu {\meanv{U}}''.
\label{equ:fpi}
\end{equation}
Assuming that there is no turbulence ($-\overline{\bm{u} \cdot
  \bm\nabla\bm{u}}=0$) and no large-scale flows ($\meanv{U}=0$), apart
from the imposed shear $\meanv{U}^{(0)}=(0,Sx,0)$, one finds the
following equations
\begin{eqnarray}
\pd_x P &=& 0, \label{equ:U01} \\
\pd_z P -2\rho\,\Omega_0 S x &=& 0. \label{equ:U02}
\end{eqnarray}
We see that Eq.~(\ref{equ:U01}) indicates that the pressure $P$ is
independent of $x$. The other equation (\ref{equ:U02}), however, gives
an explicit $x$-dependence for $P$. Therefore, we can conclude that
unless there exists some sort of large-scale forcing to cancel the
pressure gradient, there can be no equilibrium for this system. As
noted by Leprovost \& Kim (\cite{LK08b}), in the real physical
situations phenomena like the thermal winds can balance the system.


\begin{thebibliography}{}

\bibitem[1991]{BH91} Balbus, S. A. \& Hawley, J. F. 1991, \apj, 376,
  214

\bibitem[1998]{BH98} Balbus, S. A. \& Hawley, J. F. 1998, RvMP, 70, 1

\bibitem[2002]{BF02} Blackman, E. G. \& Field, G. B. 2002, PhRvL, 89,
  265007

\bibitem[2003]{BF03}
  Blackman, E. G. \& Field, G. B.,
  Phys.\ Fluids {\bf 15} (2003) L73

\bibitem[1992]{BMT92} Brandenburg, A., Moss, D. \& Tuominen, I. 1992,
  A\&A, 265, 328

\bibitem[2002]{BranDobler2002} Brandenburg, A. \& Dobler, W. 2002,
  Comp. Phys. Comm., 147, 471

\bibitem[2003]{Brandenburg2003} Brandenburg, A. 2003, in
  \emph{Advances in nonlinear dynamos (The Fluid Mechanics of
    Astrophysics and Geophysics, Vol. 9)}, ed. A. Ferriz-Mas \&
  M. N$\acute{\rm u}\tilde{\rm n}$ez, Taylor \& Francis, London and
  New York, 269
  
\bibitem[2004]{BKM04} Brandenburg, A., K\"apyl\"a, P. J., Mohammed,
  A. 2004, Phys. of Fluids, 16, 1020

\bibitem[2005]{BS05} Brandenburg, A. \& Subramanian, K. 2005, \aap,
  439, 835

\bibitem[2007]{BS07} Brandenburg, A. \& Subramanian, K. 2007, AN, 328,
  507

\bibitem[2002]{BT02} Brun, A. S. \& Toomre, J. 2002, \apj, 570, 865

\bibitem[2001]{C01} Chan, K. L. 2001, \apj, 548, 1102

\bibitem[1961]{C61} Chandrasekhar, S. 1961, \emph{Hydrodynamic and
    Hydromagnetic stability} (Oxford Univ.\ Press)

\bibitem[2003]{EKR03} Elperin, T., Kleeorin, N. \& Rogachevskii,
  I. 2003, PhRvE, 68, 016311

\bibitem[2007]{EGKR07} Elperin, T., Golubev, I., Kleeorin, N. \&
  Rogachevskii, I. 2007, PhRvE, 76, 066310

\bibitem[2007]{Fea07} Fromang, S., Papaloizou, J., Lesur, G. \&
  Heinemann, T. 2007, \aap, 476, 1123

\bibitem[2005]{GO05} Garaud, P. \& Ogilvie, G. I. 2005, JFM, 530, 145

\bibitem[2004]{KKT04} K\"apyl\"a, P. J., Korpi, M. J. \& Tuominen,
  I. 2004, A\&A, 422, 793

\bibitem[2007]{KB07} K\"apyl\"a, P. J. \& Brandenburg, A. 2007 AN,
  328, 1006

\bibitem[2008]{KB08} K\"apyl\"a, P. J. \& Brandenburg, A. 2008 A\&A,
  488, 9

\bibitem[2009]{KMB09} K\"apyl\"a, P. J., Mitra, D. \& Brandenburg,
  A. 2009, PhRvE, 79, 016302

\bibitem[1993]{KR93} Kitchatinov, L. L. \& R\"udiger, G. 1993, \aap,
  276, 96

\bibitem[1994]{KPR94} Kitchatinov, L. L., Pipin, V. V. \& R\"udiger,
  G. 1994, AN, 315, 157

\bibitem[2005]{KR05} Kitchatinov, L. L. \& R\"udiger, G. 2005,
  AN, 326, 379

\bibitem[2006]{KR06} Kleeorin, N. \& Rogachevskii, I. 2006, PhRvE, 73,
  046303

\bibitem[1974]{KR74} Krause, F. \& R\"udiger, G. 1974, AN,
  295, 93

\bibitem[1980]{KR80} Krause, F. \& R\"adler, K.-H. 1980,
  \emph{Mean-Field Magnetohydrodynamics and Dynamo Theory} (Pergamon
  Press, Oxford)

\bibitem[1993]{KRK93} K\"uker, M., R\"udiger, G. \& Kitchatinov,
  L. L. 1993, \aap, 279, 1L

\bibitem[2007]{LK07} Leprovost, N. \& Kim, E.-J. 2007, \aap, 463, 9L

\bibitem[2008a]{LK08a} Leprovost, N. \& Kim, E.-J. 2008a, PhRvE, 78,
  016301

\bibitem[2008b]{LK08b} Leprovost, N. \& Kim, E.-J. 2008b, PhRvE, 78,
  036319

\bibitem[2009]{LKKBL09} Liljestr\"om, A. J., Korpi, M. J., K\"apyl\"a,
  P. J., Brandenburg, A. \& Lyra, W. 2009, AN, 330, 91

\bibitem[2006]{Mea06} Miesch, M. S., Brun, A. S. \& Toomre, J. 2006,
  \apj, 641, 618

\bibitem[2008]{Mea08} Miesch, M. S., Brun, A. S., DeRosa, M. L.  \&
  Toomre, J. 2008, \apj, 673, 557

\bibitem[2007]{MG07} Miller, N. \& Garaud, P. 2007, AIPC, 948, 165

\bibitem[2009]{MKTB09} Mitra, D., K\"apyl\"a, P. J., Tavakol, R. \&
  Brandenburg, A. 2009, \aap, 495, 1

\bibitem[2003]{O03} Ogilvie, G. I. 2003, MNRAS, 340, 969

\bibitem[2006]{PCP06} Pessah, M. E., Chan, C.-K. \& Psaltis, D. 2006,
  PhRvL, 97, 221103

\bibitem[2008]{PCP08} Pessah, M. E., Chan, C.-K. \& Psaltis, D. 2008,
  \mnras, 383, 683

\bibitem[1993]{Pea93} Pulkkinen, P., Tuominen, I., Brandenburg, A.,
  Nordlund, \AA. \& Stein, R. F. 1993, A\&A, 267, 265

\bibitem[2005]{R05} Rempel, M. 2005, \apj, 622, 1320

\bibitem[2001]{RC01} Robinson, F. J. \& Chan, K. L. 2001, \mnras, 321,
  723

\bibitem[1980]{R80} R\"udiger, G. 1980, GAFD, 16, 239

\bibitem[1982]{R82} R\"udiger, G. 1982, AN, 303, 293

\bibitem[1989]{R89} R\"udiger, G. 1989, \emph{Differential Rotation
    and Stellar Convection: Sun and Solar-type Stars} (Akademie
  Verlag, Berlin)

\bibitem[2005]{REZ05} R\"udiger, G. Egorov, P. \& Ziegler, U. 2005,
  AN, 326, 315

\bibitem[2005]{Shea05} Schrinner, M., R\"adler, K.-H., Schmitt, D.,
  et al. 2005, AN, 326, 245

\bibitem[2007]{Shea07} Schrinner, M., R\"adler, K.-H., Schmitt, D.,
  et al. 2007, GAFD, 101, 81

\bibitem[1973]{SS73} Shakura, N. I. \& Sunyaev, R. A. 1973, \aap, 24,
  337

\bibitem[1960]{SV60} Spiegel, E. A. \& Veronis, G. 1960, \apj, 131,
  442

\bibitem[2003]{Yea03} Yousef, T. A., Brandenburg, A.
  \& R\"udiger, G. 2003, \aap, 411, 321

\bibitem[2008a]{Y08a} Yousef, T.A., Heinemann, T., Schekochihin,
  A.A. et al.\ 2008a, PhRvL, 100, 184501

\bibitem[2008b]{Y08b} Yousef, T.A., Heinemann, T., Rincon, F. et al.\
  2008b, AN, 329, 737

\bibitem[1959]{V59} Velikhov, E.P. 1959, Sov. Phys. JETP~36, 1398

\end{thebibliography}
\end{document}